\documentclass[12pt]{article}
\usepackage[utf8]{inputenc}
\usepackage{comment}
\usepackage{cite}
\usepackage[absolute]{textpos}
\input epsf.sty
\usepackage{cite}
\usepackage[margin=2cm]{geometry}
\usepackage{verbatim}         	
\usepackage{graphicx}				
\usepackage{cite}	
\usepackage{hyperref}
\usepackage{amssymb}
\usepackage{amsmath}
\usepackage[dvipsnames]{xcolor}
\newcommand{\be}{\begin{eqnarray}\displaystyle}
\newcommand{\ee}{\end{eqnarray}}
\newcommand{\nn}{\nonumber}

\newcommand{\f}{\frac}
\newcommand{\C}{\textbf{c}}
\newcommand{\Oo}{\mathcal{O}}
\newcommand{\p}{\partial}
\newcommand{\q}{\bar{q}}
\newcommand{\A}{\hat{A}}

\hypersetup{
  colorlinks   = true, 
  urlcolor     = blue, 
  linkcolor    = blue, 
  citecolor   = red  
}

\title{Asymptotic conservation law with Feynman boundary condition}
\author{}
\date{}					
\begin{document}	
\begin{textblock}{5}(6,1)
 \color{red}\Large $||$ Sri Sainath $||$
\end{textblock}
\color{black}
\maketitle

\centerline{\large {  Sayali Atul Bhatkar,  }}

\vspace*{4.0ex}

\centerline{\large \it Indian Institute of Science Education and Research,}
\centerline{\large \it  Homi Bhabha Rd, Pashan, Pune 411 008, India.}

\vspace*{1.0ex}
\centerline{\small E-mail: sayali014@gmail.com.}
\vspace{3cm}
\textbf{Abstract}\\

Recently it was shown that classical electromagnetism admits new asymptotic conservation laws \cite{2007.03627}. 
In this paper we derive the analogue of the first of these asymptotic conservation laws upon imposing Feynman boundary condition on the radiative field. We also show that the Feynman solution at $\mathcal{O}(e^3)$ contains purely imaginary modes falling off as $\lbrace \frac{\log u}{u^nr}, n\geq 0\rbrace$ which are absent in the retarded solution. The $\log u$ mode has also appeared in \cite{1903.09133,1912.10229} and violates the Ashtekar-Struebel fall offs for the radiative field\cite{AS}. We expect that new $\frac{(\log u)^m}{u^mr}$-modes would appear in the Feynman solution at order $\mathcal{O}(e^{2m+1})$. Thus, all the other modes are expected to preserve the Ashtekar-Struebel fall offs.\\\\

\newpage

\tableofcontents
\vspace{2cm}
\section{Introduction}
Soft theorems in gravitational and gauge theories\cite{soft0,soft1,soft2,soft3,soft4,soft5} are related to asymptotic conservation laws. In the classical theory these conservation laws take following form :
\be
Q^+(\hat{x})\ |\ _{\mathcal{I}^+_-}\ \  = \ \ Q^-(-\hat{x})\ |\ _{\mathcal{I}^-_+}.\label{cons}
\ee
Here, the future charge $Q^+$ is defined at $\mathcal{I}^+_-$ i.e. the $u\rightarrow-\infty$ sphere of  the future null infinity denoted by $\mathcal{I}^+$ and $u=t-r$ is its null generator. Similarly, the past charge $Q^-$ is defined at $\mathcal{I}^-_+$ which is the $v\rightarrow\infty$ sphere of the past null infinity $\mathcal{I}^-$  and $v=t+r$ is the null generator of $\mathcal{I}^-$.
In \cite{qed1,qed2,qed3, lectures, lead asym}, the authors discussed the symmetry underlying the leading soft photon theorem. They also showed that the corresponding charges obey an asymptotic conservation law as given in \eqref{cons}. This line of investigation was further extended in \cite{infinite asym}; it was shown that classical radiative field at $\mathcal{O}(e)$ admits an infinite number of conservation laws. They also provided evidence that these conservation laws are related to the infinite number of  tree level soft theorems proved in \cite{Hamada Shiu, soft inf}\footnote{The subleading terms admit corrections in presence of non-minimal couplings \cite{non uni}.}. Thus, tree level soft theorems in QED can be related to asymptotic conservation laws. 

In four spacetime dimensions, soft theorems admit non trivial loop corrections beyond the leading order term  \cite{loop1,loop2,loop3}. In \cite{1808.03288,Bis}, the authors derived the subleading soft theorems for loop amplitudes; these subleading terms involve logarithms of the soft energy and are absent in the tree level analysis. These terms are closely tied to the long range forces present in four spacetime dimensions. It is natural to ask if the loop level soft theorems are related to new asymptotic conservation laws. Accordingly in \cite{2007.03627}, we incorporated the effect of long range electromagnetic force on the scattering particles and showed that there exists a new family of conservation laws for classical electromagnetism. It is expected that these conservations laws are related to the loop level soft theorems.

The first of these conservation laws relates the coefficient of the $\frac{\log r}{r^2}$ mode at the past to the coefficient of the $\frac{\log u}{r^2}$ mode at the future ($u$ is the retarded time). This asymmetry is due to the use of retarded boundary condition. The radiation travels to the future and we have a radiative $\frac{\log u}{r^2}$ mode at the future while there exists a coulombic $\frac{\log r}{r^2}$ mode at the past. This is expected to change in the quantum theory due to the use of Feynman boundary condition. 

In this paper, our aim is to derive the analogue of this conservation law after imposing Feynman boundary condition on the radiative field. We show that the logarithmic modes obey the asymptotic conservation equation given in \eqref{1loop} by calculating the radiative field generated by scattering of $n$ charged point particles. 
An interesting aspect is that the Feynman solution contains new modes that are absent in the classical solution obeying retarded boundary condition. We expect that presence of such modes is a generic feature of QED processes. In this paper, we also aim to study the structure of these 'quantum' modes using this simplified setup. Though we obtain the explicit expression for these modes in this toy example of scattering of $n$ charged particles we will argue that many features of these modes are universal.\\\\
\textbf{Outline of this paper}\\\\
In this paper our plan is to calculate the asymptotic radiative field generated by a general scattering event involving $n$ number of charged point particles using Feynman propagator. Although this is not a physical problem, the so obtained Feynman radiative solution is useful to illustrate many interesting properties of the quantum gauge field. We will use this toy example to study certain universal aspects of the quantum gauge field. We work perturbatively in the coupling '$e$' as well as in asymptotic parameters $1/r$ (or $1/t$). 

In a scattering process some $n$ number of charged particles come in to interact and eventually move away from each other. We can divide the entire spacetime into two parts : a bulk region which is a sphere of radius $R$ around the origin $r=0$ such that the non trivial interaction between the particles takes place within this sphere. In this region, the particles in general will move on complicated trajectories depending on short range forces present between them. The second region is the asymptotic region $r>R$ in which we can completely ignore the short range forces. In the asymptotic region, we need to include the effect of the long range electromagnetic interaction that starts at $\mathcal{O}(\frac{1}{r^2})$. This is done perturbatively in $e$ and $1/r$. 

 In section 2, we  start by reviewing the asymptotic expansion of radiative field generated by above scattering event imposing retarded boundary condition.  In section 3, we obtain the radiative field generated by the same scattering event using Feynman propagator. 
In section 4, we discuss some of the 'quantum' modes present in the Feynman solution and their relation to soft modes of the quantum gauge field ($\omega\rightarrow 0$ limit) by studying the insertion of the quantum operator in an S-matrix element. We study the effect of long range electromagnetic force on the scattering particles in section 5. In section 6, we derive resultant radiative field at $\mathcal{O}(e^3)$ using Feynman propagator and discuss the new modes in the asymptotic field that arise as a result of the long range interactions between the scattering particles. We also discuss new modes that would possibly appear in the Feynman solution beyond $\Oo(e^3)$. In section 7, we discuss asymptotic conservation equation obeyed by the $\mathcal{O}(e^3)$ logarithmic modes of the Feynman radiative solution. 
Finally we summarise our results in section 8.


\section{Radiative field  with Retarded propagator}\label{2}


In this section we will obtain the asymptotic expansion of the radiative field near future null infinity using retarded propagator. The flat metric takes following form in retarded co-ordinate system ($u=t-r$) :
\be 
ds^2 = -du^2 - 2dudr + r^2\ 2\gamma_{z\bar{z}}\ dz d\bar{z}; \ \ \gamma_{z\bar{z}} = \frac{2}{(1+z\bar{z})^2}.\nn
\ee
$\mathcal{I}^+$ corresponds to the limit $r\rightarrow\infty$ with $u$ finite. We use $\hat{x}$ or $(z,\bar{z})$ interchangeably to describe points on $S^2$. We will often use following parametrisation of a 4 dimensional spacetime point (Greek indices will be used to denote 4d cartesian components) :
\be x^\mu = rq^\mu + u t^\mu,\ \ \  q^\mu=(1,\hat{x}), \ \ \  t^\mu=(1,\vec{0}).\label{q}\ee
Hence $q^\mu$ is a null vector. 

In Lorenz gauge, the radiation can be obtained from the equation $\Box A_\mu=-j_\mu$. Using the retarded propagator, we get :
\begin{align}
A_\sigma(x)
&=\frac{1}{2\pi }\int d^4x'\ \delta_+([x-x']^2)\ j_\sigma(x')\ .
\end{align} 
The subscript '+' indicates that we have choose the retarded root of the $\delta$-function constraint i.e. $t>t'$. The retarded root is given by 
\begin{align}
t'_0&=t-|\vec{x}-\vec{x}'|\label{retroot}.
\end{align}
The form of $t'_0$ at large $r$ is $t'_0=u+\mathcal{O}(\frac{1}{r})$. Thus the field $A_\sigma(r,u,\hat{x})$ at large $r$ gets contribution from $t'\sim u$. The bulk region corresponds to $|r'|<R$ or $|t'|<R$ (as $c=1$) and it contributes to $A_\sigma$ at $|u|<R$. It is a characteristic of the retarded propagator that the asymptotic field at large $u$ does not get contribution from the bulk region $|t'|<R$. Thus we can focus only on the asymptotic ($t'>R$) trajectories. 

Let us write down the form of the source current that describes our scattering event. Denoting the respective incoming velocities by $V_i^\mu$, charges by $e_i$ and masses by $m_i$ (for $ i=1\cdots n$). Restricting ourselves to the leading order in coupling $e$, we ignore the effect of long range electromagetic interactions on the asymptotic trajectories. Thus an incoming particle has the trajectory  :
$$ x^\mu_i= [V_i^\mu \tau + d_i]\Theta(-T-\tau).$$
$\tau$ is an affine parameter, here $T$ denotes the value of $\tau$ such that $r(-T)=R$ hence the short range forces can be ignored for $\tau<-T$. Similarly let us denote the asymptotic outgoing velocities of the particles by $V_j^\mu$, charges $e_j$ and masses $m_j$ (for $j=n+1\cdots 2n$), an outgoing particle has the trajectory :
$$ x^\mu_j= [V_j^\mu \tau + d_j]\Theta(\tau-T).$$
For outgoing particles $T$ denotes the value of $\tau$ such that $r(T)=R$ hence the short range forces can be ignored for $\tau >T$. The current is given by summing over all particles that participate in the scattering. The asymptotic part of this current at $\Oo(e)$ can be written down as :
\be j^{\text{asym}}_\sigma(x') &=& \int d\tau\Big[\sum_{i=n+1}^{2n} e_i V_{i\sigma}\  \delta^4(x'-x_i)\ \Theta(\tau-T)+\sum^{n}_{i= 1} e_i V_{i\sigma}\  \delta^4(x'-x_i)\ \Theta(-T-\tau)\Big]. \label{j}\ee

The asymptotic radiative field generated by the scattering process can be obtained as follows
\begin{align}
A_\sigma(x)&=\frac{1}{2\pi }\int d^4x'\ \delta_+([x-x']^2)\ j^{\text{asym}}_\sigma(x')\ ,\nn\\
&=\sum_{i=n+1}^{2n}\frac{1}{4\pi }  \frac{e_iV_{i\sigma}\ \Theta(\tau^+_0-T) }{\sqrt{(V_i.x-V_i.d_i)^2+(x-d_i)^2}}  +\sum_{i=1}^{n}\frac{1}{4\pi }  \frac{e_iV_{i\sigma}\ \Theta(-\tau^+_0-T) }{\sqrt{(V_i.x-V_i.d_i)^2+(x-d_i)^2}}\label{A0} \ .
\end{align}
Here, $\tau_0^+=-(V_i.x-V_i.d_i)-\sqrt{(V_i.x-V_i.d_i)^2+(x-d_i)^2}$ is the retarded root. To expand the radiative field around $\mathcal{I}^+$ let us take the limit $r\rightarrow\infty$ with $u$ finite in eq.\eqref{A0}. Using $\tau^+_0= \frac{u+q.d_i}{|q.V_i|}+\mathcal{O}(\frac{1}{r})$ in \eqref{A0} :
\begin{align}
A_\sigma(x)|_{\mathcal{I}^+}
&=-\frac{1}{4\pi r}\Big[\sum_{i=n+1}^{2n}  \frac{e_iV_{i\sigma}}{V_i.q} \Theta(u+q.d_i-T) +\sum_{i=1}^{n} \frac{e_iV_{i\sigma}}{V_i.q}  \Theta(-u-q.d_i-T) + ...\Big]+\mathcal{O}(\frac{1}{r^2})\ . \label{AA1}
\end{align}
At large values of $u$, we see that the $\frac{1}{r}$-term goes like $u^0$. This mode is resposible for the so called memory effect \cite{mem1,mem2,mem3}. We use '$...$' to denote $u$-fall offs that are faster than any (negative) power law behaviour. Let us rewrite \eqref{AA1} a bit succinitly :
\begin{align}
A_\mu(x)=\frac{1}{4\pi r} \ A_\mu^{[1,0]}(\hat{x}) \ u^0\  +\mathcal{O}(\frac{1}{r^2})\ ,\ \ \ | u|\rightarrow\infty.\nn
\end{align}

Next we study the $\mathcal{O}(\frac{1}{r^2})$ term. We need to expand the $\frac{1 }{\sqrt{(V_i.x-V_i.d_i)^2+(x-d_i)^2}} $ factor in \eqref{A0} to $\mathcal{O}(\frac{1}{r^2})$. We get 
\begin{align}
A_\sigma^{[2]}(u,\hat{x})
&=-\frac{1}{4\pi }\ \sum_{i=n+1}^{2n}  \frac{e_iV_{i\sigma}}{(V_i.q)^2} \Big[ u\ [V_i^0+\frac{1}{V_i.q}]+\ V_i.d_i+\frac{d_i.q}{V_i.q} \ \Big]\Theta(u+q.d_i-T)+\text{  in  }\ .\nn
\end{align}
$A_\sigma^{[2]}$ denotes the coefficient of the $\mathcal{O}(\frac{1}{r^2})$ term in the radiative field. In above expression we have not written down the contribution of incoming particles to avoid clutter. It takes the same form as the contribution of the outgoing particles but comes with a factor of $\Theta(-u)$. From above expression we see that the leading term at $|u|\rightarrow\infty$ is $\mathcal{O}(u)$, it is followed by a $\mathcal{O}(u^0)$ term and then there are highly suppressed terms like $\delta(u)$ which we are not keeping track of. Thus, analogous to \eqref{AA1} we get 
\begin{align}
A_\mu(x)=\frac{1}{4\pi r} \ \Big[\ A_\mu^{[1,0]}(\hat{x})\ u^0\ + \frac{1}{ r} \ \big[\ A_\mu^{[2,-1]}(\hat{x}) \ u +  A_\mu^{[2,0]}(\hat{x})\ u^0\ \big]\  \Big] +\mathcal{O}(\frac{1}{r^3})\ ,\ \ \  | u|\rightarrow\infty.\nn
\end{align}

This can be extended to higher orders in $\frac{1}{r}$ and we see that the asymptotic expansion of the radiative field around $\mathcal{I}^+$ takes following form 
\begin{align}
A^{\text{ret}}_\mu(x)|_{\mathcal{I}^+}=\frac{1}{4\pi } \sum_{\substack{m=0,n=1\\ m<n}}^\infty  [A_\mu^{[n,-m]}(\hat{x})]^{\text{ret}} \ \frac{u^m}{ r^n} +...\ , \ \ \ | u|\rightarrow\infty.\label{AAn}
\end{align}
Here '...' denote the terms that fall off faster than any power law. It should be noted that above expression is valid at $\mathcal{O}(e)$ as it was obtained ignoring the asymptotic electromagnetic force present between the scattering particles. We have used a superscript 'ret' to recall that \eqref{AAn} corresponds to retarded boundary condition. Next we will study the analogue of \eqref{AAn} after imposing Feynman boundary condition.

\section{Radiative field with Feynman propagator}\label{3}
In this section we will obtain the radiative field produced by scattering of $n$ charged particles upon imposing Feynman boundary condition. This is not a physical problem but this simple setup helps us to study certain aspects of the quantum gauge field. 
As the Feynman propagator has an imaginary piece, the resultant radiative field also has an imaginary piece. As we shall see in section \ref{4}, the presence of these imaginary modes is in fact tied to the soft modes of the quantum field.

As discussed in \eqref{j}, the scattering event is described by following current
\be j^{\text{asym}}_\sigma(x') = \int d\tau\Big[\sum_{i=n+1}^{2n} e_i V_{i\sigma}\  \delta^4(x'-x_i)\ \Theta(\tau-T)+\sum^{n}_{i= 1} e_i V_{i\sigma}\  \delta^4(x'-x_i)\ \Theta(-\tau-T)\Big].\label{jj}\ee
Here, we have labelled the incoming particles by $i$ running from 1 to $n$ and outgoing particles by $i$ running from $n+1$ to $2n$. We do not have an explicit form of the bulk trajectories i.e. for $|\tau|<T$. The radiative field is given by 
\begin{align}
A_\sigma(x)
&=\ \int d^4x'\ G(x,x')\ j_\sigma(x')\ ,
\end{align}
using the usual momentum representation of the Feynman propagator we have, 
\be G(x,x')
&=&\  \int \frac{d^4p}{(2\pi)^{4}}\   \frac{e^{ip.(x-x')}}{p^2-i\epsilon}.\nn
\ee
We can perform the momentum integral to obtain the form of the propagator in the position space. 
\be
G(x;x')
&=& \frac{1}{4\pi^2 } \Big[\pi\delta_+(\ (x-x')^2)+\pi\delta_-(\ (x-x')^2)+\frac{i}{(x-x')^2}\Big]\label{fp}.
\ee
See Appendix \ref{FP} for derivation of above expression. The subscript '+' denotes the retarded root of the $\delta$-function constraint i.e. $t>t'$, while the subscript '-' denotes the advanced root of the $\delta$-function constraint i.e. $t'>t$. 

Since, the first term in above expression is proportional to the retarded propagator, the electromagnetic field generated by this term is similar to the one obtained in the previous section. We denote the field generated by the first term in \eqref{fp} by superscript '+'. We have discussed in the previous section using \eqref{retroot} that the asymptotic field gets contribution only from the asymptotic part of the current.  So using \eqref{jj}, we get
\begin{align}
A^+_\sigma(x)
&=\frac{1}{8\pi }\sum_{i=n+1}^{2n}  \frac{e_iV_{i\sigma}\ \Theta(\tau^+_0-T) }{\sqrt{(V_i.x-V_i.d_i)^2+(x-d_i)^2}}  +\frac{1}{8\pi }\sum_{i=1}^{n}  \frac{e_iV_{i\sigma}\ \Theta(-\tau^+_0-T) }{\sqrt{(V_i.x-V_i.d_i)^2+(x-d_i)^2}} \ .\label{+}
\end{align}
Here the retarded root is given by $\tau^+_0=-(V_i.x-V_i.d_i)-\sqrt{(V_i.x-V_i.d_i)^2+(x-d_i)^2}.$ The asymptotic expansion of above expression around $\mathcal{I}^+$ is similar to \eqref{AAn}.
\begin{align}
A^+_\mu(x)|_{\mathcal{I}^+}=\frac{1}{8\pi } \sum_{\substack{m=0,n=1\\ m<n}}^\infty  [A_\mu^{[n,-m]}(\hat{x})]^+  \ \frac{u^m}{ r^n} +...\ ,\label{+e}
\end{align}
 where '...' denote the terms that fall off faster than any power law. 

The second term in \eqref{fp} is proportional to the advanced propagator, we will denote the field generated by this term by the superscript '$-$'. Due to the reasons similar to the retarded case, this term also does not get any contribution from the bulk current and it suffices to use \eqref{jj}. We have
\begin{align}
A^-_\sigma(x)
&=\frac{1}{8\pi }\sum_{i=n+1}^{2n}  \frac{e_iV_{i\sigma}\ \Theta(\tau^-_0-T) }{\sqrt{(V_i.x-V_i.d_i)^2+(x-d_i)^2}}  +\frac{1}{8\pi }\sum_{i=1}^{n}  \frac{e_iV_{i\sigma}\ \Theta(-\tau^-_0-T) }{\sqrt{(V_i.x-V_i.d_i)^2+(x-d_i)^2}} \ .\label{-}
\end{align}
Here the advanced root is given by $\tau^-_0=-(V_i.x-V_i.d_i)+\sqrt{(V_i.x-V_i.d_i)^2+(x-d_i)^2}.$ We can expand above expression around $\mathcal{I}^+$, an important point to notice here is that $ \tau_0^-|_{\mathcal{I}^+}= 2r{|q.V_i|}+\mathcal{O}(r^0).$ Substituting this value in above expression, the step function with the incoming particles goes like $\Theta(-r)$ hence the contribution of the incoming particles in above expression goes to 0. The asymptotic expansion takes following form, 
\begin{align}
A^-_\mu(x)|_{\mathcal{I}^+}=\frac{1}{8\pi } \sum_{\substack{m=0,n=1\\ m<n}}^\infty  [A_\mu^{[n,-m]}(\hat{x})]^-  \ \frac{u^m}{ r^n} +...\ .\label{-e}
\end{align}
The coefficients $[A]^-$ should be contrasted with $[A]^+$ in \eqref{+e}. $[A]^-$ are same throughout $\mathcal{I}^+$ from $u\rightarrow-\infty$ to $u\rightarrow\infty$ while the coefficients $[A]^+$ in the retarded solution take differents values at $u\rightarrow\pm\infty$ respectively.

Finally we turn to the contribution from the third term in \eqref{fp} i.e. from $\frac{i}{4\pi^2(x-x')^2}$, we denote it by superscript '$^*$'. This term gets contribution from all of spacetime including the bulk. 
\begin{align}
A^{*}_\mu(x)
&=\frac{i}{4\pi^2 }\ \int d^4x'\frac{j_{\mu}(x')}{(x-x')^2}.\nn
\end{align}
We will use \eqref{jj} for the asymptotic part of the current, but the explicit form of the bulk current is not available. Using \eqref{jj}, we get
\begin{align}
A^{*}_\sigma(x)
&=\frac{i}{4\pi^2 }\Big[\int_T^\infty d\tau \sum_{i=n}^{2n} \frac{e_i V_{i\sigma}}{(x-V_i\tau-d_i)^2}+\int^{-T}_{-\infty} d\tau \sum_{i=1}^{n} \frac{e_i V_{i\sigma}}{(x-V_i\tau-d_i)^2}+ \int_{r'<R} d^4x'\frac{j_{\sigma}(x')}{(x-x')^2}\Big].\label{b2}
\end{align}
First we focus on the  asymptotic contribution. 
\begin{align}
A^{*\text{asym}}_\sigma(x)
&=-\frac{i}{4\pi^2 }\ \Big[\int_T^\infty d\tau \sum_{j=n}^{2n} \frac{e_j V_{j\sigma}}{(\tau-\tau_0^+)(\tau-\tau_0^-)}+\int^{-T}_{-\infty} d\tau \sum_{j=1}^{n} \frac{e_j V_{j\sigma}}{(\tau-\tau_0^+)(\tau-\tau_0^-)}\Big].\nn
\end{align}
$\tau^\pm_0$ are the solutions to the equation $(x-V_i\tau-d_i)^2=0$ and the expressions are given in \eqref{tau00}. The integral involving the outgoing particles has a divergence at the upper limit. Let us regulate it with an IR cutoff  '$R$'. Similarly we regulate the second integral with a cutoff '$-R$' to get
\begin{align}
A^{*\text{asym}}_\sigma(x)
&=\frac{i}{4\pi^2 }\sum_{j=n+1}^{2n}\frac{e_j V_{j\sigma}}{\tau_0^--\tau_0^+}\Big[\log\frac{R-\tau_0^+}{T-\tau_0^+}-\log\frac{R-\tau_0^-}{T-\tau_0^-}\Big]-\frac{i}{4\pi^2 }\sum_{j=1}^{n}\frac{e_j V_{j\sigma}}{\tau_0^--\tau_0^+}\Big[\log\frac{R+\tau_0^+}{T+\tau_0^+}-\log\frac{R+\tau_0^-}{T+\tau_0^-}\Big]\ .\nn
\end{align}
All the quantities appearing in the argument of the log function come with a modulus sign which we do not write down explicitly. We expand the square brackets in the limit $R\rightarrow\infty$ and see that the divergent pieces cancel. Hence the final expression is finite, we get
\begin{align}
&A^{*\text{asym}}_\sigma(x)=\frac{i}{4\pi^2 }\ \sum_{j=n+1}^{2n}\frac{e_j V_{j\sigma}}{\tau_0^--\tau_0^+}\ \log\frac{\tau_0^--T}{\tau_0^+-T}-\frac{i}{4\pi^2 }\ \sum_{j=1}^{n}\frac{e_j V_{j\sigma}}{\tau_0^--\tau_0^+}\ \log\frac{\tau_0^-+T}{\tau_0^++T}.\nn
\end{align}
We will use \eqref{tau00} to substitute for $\tau_0^--\tau_0^+$ and also rewrite above expression in a succinit form 
\begin{align}
A^{*\text{asym}}_\sigma(x)&=\frac{i}{8\pi^2 }\ \sum_{j=1}^{2n}\frac{\eta_je_j V_{j\sigma}}{\sqrt{(V_j.x-V_j.d_j)^2+(x-d_j)^2}}\ \log\frac{\tau_0^--\eta_jT}{\tau_0^+-\eta_jT}.\label{*}
\end{align}
Here $\eta_j=1(-1)$ for outgoing (incoming) particles. Next we will find the asymptotic expansion of above expression. Using \eqref{tau0f} we have $$\tau_0^+|_{\mathcal{I}^+}= \frac{u+q.d_i}{|q.V_i|}+\mathcal{O}(\frac{1}{r}),\ \ \tau_0^-|_{\mathcal{I}^+}= 2r{|q.V_i|}+\mathcal{O}(r^0).$$
Thus we get
$$\Big[\log\frac{\tau_0^--T}{\tau_0^+-T}\Big]_{\mathcal{I}^+}=\ \log\frac{r}{u}+\mathcal{O}(1).$$
We find that there are logarithmic modes in the radiative field.  This is an interesting result as we will discuss at the end of this calculation. Let us write down the full asymptotic expansion of $A^{*\text{asym}}_\sigma$. Using \eqref{tau00}, it is seen that $$\log \tau_0^-|_{\mathcal{I}^+}\sim\ \log r\ + \sum_{\substack{m,n=0,\\ m\leq n.}}^\infty \frac{u^m}{r^n}.$$
Similarly $$\log \tau_0^+|_{\mathcal{I}^+}\sim\  \log u\ + \sum_ {\substack{n=0,\\ m=-\infty,\\m\leq n.}}^\infty \frac{u^m}{r^n}.$$
We will write down the expansion for $A^{*\text{asym}}_\sigma$ by substituting above expressions in \eqref{*}.
\begin{align}
A^{*\text{asym}}_\sigma(x)
&\sim\log\frac{ u}{r} \sum_{\substack {m=0,n=1\\m< n}}^\infty \frac{u^m}{r^n}\ + \sum_{\substack {n=1,\\ m=-\infty,\\m< n.}}^\infty \frac{u^m}{r^n}.
\end{align}

Next we turn to the bulk contribution i.e. the $r'<R$ term in \eqref{b2}. We do not have the explicit expression of the bulk current.
\begin{align}
 A^{*\text{bulk}}_\sigma(x)
&=\frac{i}{4\pi^2 }\  \int_{r'<R} d^4x'\ j_{\sigma}(x')\ \frac{1}{(x-x')^2}\ .
\end{align}
Let us estimate the contribution of this integral around $\mathcal{I}^+$. $\frac{1}{(x-x')^2}|_{\mathcal{I}^+}\sim \frac{1}{r}\sum_{n=1}^\infty\frac{1}{u^n}+\mathcal{O}(\frac{1}{r^2})$.
Similarly the full asymptotic expansion of $A^{*\text{bulk}}_\sigma(x)$ around the future null infinity can be written down.
\be A^{*\text{bulk}}_\sigma(x)|_{\mathcal{I}^+}\sim \  \sum_{\substack{n=1,\\ m=-\infty,\\ m<n-1.}}^\infty\frac{u^m}{r^n} \ .\label{*bulk} \ee

Finally we write down the Feynman solution using \eqref{+},\eqref{-} and \eqref{*} :
\begin{align}
A_\sigma(x)&=A^{+}_\sigma(x)+A^{-}_\sigma(x)+A^{*}_\sigma(x)\nn\\
&=\frac{1}{8\pi }\sum_{i=1}^{2n}  \frac{e_iV_{i\sigma}\ \Theta(\eta_i\tau^+_0-T) }{\sqrt{(V_i.x-V_i.d_i)^2+(x-d_i)^2}}  +\frac{1}{8\pi }\sum_{i=n+1}^{2n}  \frac{e_iV_{i\sigma}\ \Theta(\tau^-_0-T) }{\sqrt{(V_i.x-V_i.d_i)^2+(x-d_i)^2}}  \ \nn\\
&+\frac{i}{8\pi^2 }\ \sum_{j=1}^{2n}\frac{\eta_je_j V_{j\sigma}}{\sqrt{(V_j.x-V_j.d_j)^2+(x-d_j)^2}}\ \log\frac{\tau_0^--\eta_jT}{\tau_0^+-\eta_jT}+A^{*\text{bulk}}_\sigma(x).\label{Afeyn}
\end{align}
As before, $\eta_j=1(-1)$ for outgoing (incoming) particles. We do not have an explicit form for $A^{*\text{bulk}}_\sigma$. It will depend on the details of the scattering process and short ranges forces present between the particles. We are not interested in these non-universal terms. Nonetheless we will see in the next section that $A^{*\text{bulk}}_\sigma$ has some universal modes and we will use a trick to calculate them.

We have already studied the asymptotic expansion of above solution (including $A^{*\text{bulk}}_\sigma$) around $\mathcal{I}^+$. Let us comment on some important differences between the Feynman solution and the retarded solution. The leading order term of \eqref{Afeyn} is $\Oo(\frac{\log r}{r})$. If we study \eqref{AAn}, we see that such kind of modes are completely absent in the retarded solution! The retarded solution discussed in \eqref{AAn} starts at $\Oo(\frac{1}{r})$. The $\frac{1}{r}$- component of the Feynman solution given in \eqref{Afeyn} takes following form 
\begin{align}
A_\sigma(x)|_{\mathcal{I}^+}\sim\frac{1}{4\pi r} \ [\ \log u + u^0 + \sum_{n=1}^\infty\frac{1}{u^n} + ...\ ] \    .\label{AFP1}
\end{align}
Above expression should be contrasted with \eqref{AA1}. The $\log u$ mode is absent in the classical field. It is very important to note that this mode violates the Ashtekar-Struebel fall offs for the radiative field\cite{AS} and needs to be studies further. 
We expect that the presence of a log $u$ mode is a general feature of QED. One way to prove this statement is to relate the log $u$ mode to the universal leading soft mode as done in \cite{1903.09133,1912.10229}. We will discuss this derivation in the next section. 

Let us turn to the coefficient of the $\frac{u^0}{r}$ mode. 
\begin{align}
A_\sigma^{[1,0]}(\hat{x})
&= -\frac{1}{8\pi }\sum_{i=n+1}^{2n}  \frac{e_iV_{i\sigma}}{V_i.q} \Theta(u+q.d_i-T) -\frac{1}{8\pi } \sum_{i=1}^{n} \frac{e_iV_{i\sigma}}{V_i.q}  \Theta(-u-q.d_i-T)\nn\\
&-\frac{1}{8\pi }\sum_{i=n+1}^{2n}  \frac{e_iV_{i\sigma}}{V_i.q}-\frac{i}{4\pi^2 }\ \sum_{j=1}^{2n}[\eta_je_j\log |q.V_i|\ + \eta_je_j\log \sqrt{2}]\ \frac{V_{j\sigma}}{q.V_j}.\label{c1}
\end{align}
It is well known that the $\frac{u^0}{r}$ mode in the gauge field is related to the memory effect and is expected to be controlled by the leading soft factor \cite{mem1,mem2,mem3}. Apriori it might seem that \eqref{c1} is in conflict with the previous statement. But let us recall that the memory effect is proportional to the change of gauge field from $u=-\infty$ to $u=\infty$. Using \eqref{c1}, we get
\begin{align}
\Delta A_\sigma^{[1,0]} = \int_{\mathcal{I}^+} du\ \p_uA_\sigma^{[1,0]} = -\frac{1}{8\pi }\sum_{i=1}^{2n}\eta_i  \frac{e_iV_{i\sigma}}{V_i.q} .\nn
\end{align}
Thus we see that $\Delta A_\sigma$ is in fact proportional to the leading soft factor. Though there are corrections to the $\frac{u^0}{r}$ mode as seen in \eqref{c1}, these corrections do not contribute to the memory effect. As seen in \eqref{c1}, this mode also has an imaginary piece which is absent in the classical mode given in \eqref{AA1}. This 'quantum' mode has appeared in the analysis of  \cite{1912.10229}. 
Performing a co-oordinate transformation of \eqref{c1}, we calculate the radial component of the radiative field. We have 
\begin{align}
A^{1}_r(x)|_{\text{imag}}
&=-\frac{1}{4\pi^2 }\ \sum_{j=1}^{2n}\eta_je_j\log |q.V_i|\ .
\end{align}
Above expression matches with the quantum $A_r^1$ mode discussed in eqn (69) of \cite{1912.10229} in the context of massless QED coupled to gravity. This hints that the coefficent of above mode is universal. 

We will obtain the explicit for of coefficients of the log $u$ and $\frac{1}{u}$ modes in the next section. Here we wish to emphasise that these modes are not expected to give rise to memory effects as they do not contribute to $\Delta A_\sigma$. In particular the 'quantum' $\frac{1}{u}$-mode present in \eqref{AFP1} does not contribute to the tail memory effect discussed in \cite{log mem,log mem em}. It should also be noted that this $\frac{1}{u}$-mode appears at $\Oo(e)$ and is not related to the long range electromagnetic interaction. The tail memory term appears at $\Oo(e^3)$ and is a direct consequence of the long range interactions. 

We will conclude this section after writing down the asymptotic expansion of the full solution in \eqref{Afeyn}. It is given by
\begin{align}
&A_\sigma(x)|_{\mathcal{I}^+}\nn\\
&=\sum_{\substack {m=0,n=1\\m< n}}^\infty\frac{u^m}{r^n}\  A_\sigma^{[n,-m]}(\hat{x})+\log u \sum_{\substack {m=0,n=1\\m< n}}^\infty \frac{u^m}{r^n}\ L_{1\sigma}^{[n,-m]}(\hat{x}) \ +\log r \sum_{\substack {m=0,n=1\\m< n}}^\infty \frac{u^m}{r^n}\ L_{2\sigma}^{[n,-m]}(\hat{x}) \ + \sum_ {m,n=1}^\infty \frac{A_\sigma^{[n,m]}(\hat{x})}{u^mr^n}.\label{AFP}
\end{align}
Let us compare above solution with the retarded solution we have in \eqref{AAn}. The retarded solution has only $ A_\sigma^{[n,-m]}$ kind of modes. The other kind of modes present in the Feynman solution that have log behaviour or fall off as negative powers of $u$ are absent in the retarded solution (at $\mathcal{O}(e)$). 

\section{Insertion of the quantum gauge field in $S$ matrix element}\label{4}
In the last section, we derived the asymptotic expansion of Feynman solution generated by scattering of $n$ charged particles. As seen in \eqref{AFP}, it contains $\log u$ mode and $\frac{1}{u^n}$ modes that are absent in the retarded solution given in \eqref{AAn}. In this section, we will 
discuss the relation of such modes to the soft modes of the quantum field. We will insert the quantum U(1) gauge field operator and evaluate the S matrix elements and show that they match with the corresponding coefficients in \eqref{AFP}. In \cite{1903.09133,1912.10229}, the $\log u$ mode in the quantum field was derived from the leading soft mode. Extending this idea, we show that the $\frac{1}{u}$-mode is related the tree level subleading soft mode.

\subsection{The $\frac{\log u}{ r}$ mode in $A_\sigma$}
Let us start with the coefficient of the $\frac{\log u}{r}$ mode. Using the radiative field calculated in \eqref{Afeyn}, we see that this mode arises from the second line of \eqref{Afeyn}. The contribution from the first term in the second line of \eqref{Afeyn} is given by
\begin{align}
A^{[1,\log]}_\sigma(x)
&= \frac{i}{8\pi^2 }\ \sum_{j=1}^{2n}\frac{\eta_je_j V_{j\sigma}}{V_j.q}\ .\label{logu}
\end{align}
Here $A^{[1,\log]}_\sigma$ has been used to denote the coefficient of the $\frac{\log u}{r}$ mode.

In \cite{1903.09133,1912.10229}, the $\frac{\log u}{r}$ mode was derived in a different manner. 
Let us start with the quantum gauge field $\hat{A}_\mu$. In momentum space the expansion of $\hat{A}_\mu$ is given by
\be \hat{A}_\sigma({x})= \frac{1}{(2\pi)^3}\int \frac{d^3p}{2|\vec{p}|}\ [a_\sigma(p)\ e^{ip.x} +a_\sigma^\dagger(p)\ e^{-ip.x}\ ].\ee
Here, $a_\sigma(\omega,\hat{x})=\sum_{r=+,-} \epsilon^{*r}_\sigma(\hat{x}) \ a^r(\omega,\hat{x})$ such that $a^r(\omega,\hat{x})$ is identified as the annihilation operator for the respective helicity photons and $\epsilon^r_\sigma(\hat{x})$ is the polarisation vector. We will find the leading order term in above expression at $\mathcal{I}^+$ by taking the limit $r\rightarrow\infty$ with $u$ finite. Using stationary phase approximation it can be shown that the leading order term is $\Oo(\frac{1}{r})$ and its coefficient $\hat{A}^{[1]}_{\mu}(u,\hat{x})$ is given by \cite{lectures} 
\be \A^{[1]}_\sigma(u,\hat{x})= -\frac{i}{8\pi^2}\int_0^\infty d\omega\ [a_\sigma(\omega,\hat{x})\ e^{-i\omega u} +a_\sigma^\dagger(\omega,\hat{x})\ e^{i\omega u}\ ].\label{hom}\ee
 Above expression can be rewritten as 
\be \A^{[1]}_\sigma(u,\hat{x})&= &-\frac{i}{8\pi^2}\int_{-\infty}^\infty d\omega\ \tilde{A}_\sigma \ e^{-i\omega u}\ , \nn\\
\text{here}\ \ \tilde{A}_\sigma&=& [\ a_\sigma(\omega,\hat{x})\ \Theta(\omega) -a_\sigma^\dagger(-\omega,\hat{x})\Theta(-\omega)\ ] .\label{hom2}\ee

Our aim is study an $S$ matrix element of following form $<\text{out}|\A_\mu S|\text{in}>$. This element will get contribution only from the positive freqencies. We define a function $\A^{[1]+}_{\mu}(u,\hat{x})$ that has contribution from only positive frequencies i.e. 
\begin{align}
\A^{[1]+}_\mu(u,\hat{x})&=-\frac{i}{8\pi^2}\int_0^\infty d\omega\ \tilde{A}_\mu(\omega,\hat{x}) \ e^{-i\omega u}.\nn
\end{align}
In above expression let us add an imaginary part to $u$ to make the integral well defined at $\omega\rightarrow\infty$.
\begin{align}
\A^{[1]+}_\mu(u,\hat{x})&=-\frac{i}{8\pi^2}\int_0^\infty d\omega\ \tilde{A}_\mu(\omega,\hat{x}) \ e^{-i\omega (u-i\epsilon)}.\nn
\end{align}
Next we will find the behaviour of the field at large $u$ using a common trick. 
\begin{align}
\p_u\A^{[1]+}_\mu(u,\hat{x})&=-\frac{1}{8\pi^2}\int_0^\infty d\omega\ [\omega \tilde{A}_\mu(\omega,\hat{x})] \ e^{-i\omega (u-i\epsilon)},\nn\\
&=-\frac{i}{8\pi^2}\frac{1}{(u-i\epsilon)}\int_0^\infty d\omega\ [\omega \tilde{A}_\mu(\omega,\hat{x})] \ \p_{\omega}e^{-i\omega (u-i\epsilon)},\nn\\
&=-\frac{i}{8\pi^2} \frac{1}{(u-i\epsilon)}\Big[[\omega \tilde{A}_\mu(\omega,\hat{x})] e^{-i\omega (u-i\epsilon)}\Big]^\infty_0+\frac{i}{8\pi^2}\frac{1}{(u-i\epsilon)}\int_0^\infty d\omega\ \p_{\omega}[\omega \tilde{A}_\mu(\omega,\hat{x})] \ e^{-i\omega (u-i\epsilon)}.\label{d7}
\end{align}
For now we will focus on the first term (The second term will be studied in the next subsection and is subleading at large $u$). We have
\begin{align}
\p_u\A^{[1]+}_\mu(u,\hat{x})&=\frac{i}{8\pi^2}\frac{1}{(u-i\epsilon)} \ \lim_{\omega\rightarrow 0}[\omega \tilde{A}_\mu(\omega,\hat{x})] +...\nn  
\end{align}
Hence
\begin{align}
\A^{[1]+}_\mu(u,\hat{x})&=\frac{i}{8\pi^2} \ [\log u +i\pi\Theta(u)]\ \lim_{\omega\rightarrow 0}[\omega \tilde{A}_\mu(\omega,\hat{x})] +...\ .\nn
\end{align}
Thus we see that at large $u$ the field has $\log u$ and $u^0$ modes.The  coefficient of the $u^0$ term is
\begin{align}
\A^{[1,0]}_\mu(\hat{x})& =- \frac{\Theta(u)}{8\pi} \lim_{\omega\rightarrow 0^+} \ \omega \tilde{A}_\mu(\omega,\hat{x}) \nn\\
& =- \frac{\Theta(u)}{8\pi}\lim_{\omega\rightarrow 0^+} \omega a_\mu(\omega,\hat{x}) .\nn
\end{align}
Above mode is responsible for the so called memory effect \cite{mem2} and as expected its coefficient is proportional to the leading soft factor. 
Then we turn to the $\log u$ mode. We have : 
\begin{align}
\A^{[1,\log]}_\mu(\hat{x})& = \frac{i}{8\pi^2} \lim_{\omega\rightarrow 0^+} \ \omega \tilde{A}_\mu(\omega,\hat{x}) \  .\nn\\
%
& = \frac{i}{8\pi^2} \lim_{\omega\rightarrow 0^+} \omega a_\mu(\omega,\hat{x})\  .\end{align}
Thus, both the modes log $u$ and $\Theta(u)$ are related to the leading soft mode. We can evaluate the insertion of above operator using leading soft theorem.
\begin{align}
<\text{out}|\A^{[1,\log]}_\mu(\hat{x})\ S|\text{in}>\ &=\frac{i}{8\pi^2} \big[\epsilon^+_\mu\epsilon^-_\nu+\epsilon^-_\mu\epsilon^+_\nu\big] \ \sum_{j=1}^{2n}\frac{\eta_je_jV_j^\nu}{V_j.q}\  <\text{out}|S|\text{in}> ,\nn\\
& = \frac{i}{8\pi^2} \  \sum_{j=1}^{2n}\frac{\eta_je_jV_{j\mu}}{V_j.q}\  <\text{out}|S|\text{in}>  .
\end{align}
Here, $|\text{in}>\ = |1,2,...,n'>$ and $<\text{out}|=\ <n'+1,...,2n|$ aregeneric  Fock states. We see that when the quantum operator in inserted between generic states, the coefficient of the $\log u$ mode matches with our expression obtained from Feynman radiative solution in \eqref{logu}. It is clear from this derivation that the existence of the $\log u$ mode is tied to the $\frac{1}{\omega}$-mode. Since the $\frac{1}{\omega}$-mode is universal we expect that the $\log u$ mode is also universal.

\subsection{The $\frac{1}{ ur}$ mode in $A_\sigma$}
Next we turn to the $\frac{1}{ru}$ mode of the Feynman solution $A_\sigma$. This mode arises from the second line of \eqref{Afeyn}. It is interesting to note that this mode gets contribution from both bulk and asymptotic sources. Let us first write down the contribution from the first term in the second line of \eqref{Afeyn}. In this term, we substitute the expression of $\tau^+_0$ from \eqref{tau00} and use $\log (\tau_0^+-\eta_iT) = \log \frac{u}{|q.V_i|} + \frac{q.d_i}{u}+\eta_iq.V_i\frac{T}{u}+...$ , to get 
\begin{align}
A^{[1,1]}_\sigma(x)|_{\text{asym}}
&=  \frac{i}{8\pi^2 }\ \sum_{j=1}^{2n}\eta_je_j V_{j\sigma}\ [\frac{q.d_j}{V_j.q}\ +\eta_jT\ ]\ .\label{12}
\end{align}
$A^{[1,1]}_\sigma$ denotes the coefficient of the $\frac{1}{ru}$ mode of $A_\sigma$. Let us turn to the contribution from the bulk.
\begin{align}
 A^{*\text{bulk}}_\sigma(x)
&=\frac{i}{4\pi^2 }\  \int_{r'<R} d^4x'\ j_{\sigma}(x')\ \frac{1}{(x-x')^2}\nn\  .
\end{align}
We will contract it with $q^\sigma$ defined in \eqref{q} as it will allow us to exploit the conservation property of the U(1) current. This trick is similar to the one used in \cite{Sen Laddha}. Then we use $\p_\sigma'\log{(x-x')^2}=-2\frac{(x-x')_\sigma}{(x-x')^2}=-\frac{2rq_\sigma}{(x-x')^2}+\mathcal{O}(\frac{1}{r})$ to obtain
\begin{align}
q^\sigma A^{*\text{bulk}}_\sigma(x)
&=-\frac{i}{8\pi^2 r}\  \int_{r'<R} d^4x'\ j_{\sigma}(x')\ \p'^\sigma\log{(x^2-2x.x')}+\mathcal{O}(\frac{1}{r^2})\ .\nn
\end{align}
We have ignored the the $x'^2$ term in the argument of the $\log$ as we are not keeping track of $\mathcal{O}(\frac{1}{r^2})$ corrections. Integrating by parts, we are left with a boundary term 
\begin{align}
q^\sigma A^{*\text{bulk}}_\sigma(x)
&=-\frac{i}{8\pi^2 r}\  \int_{r'=R} d^4x'\  {n'^\sigma j_{\sigma}(x')}\ \log{(x^2-2x.x')}+\mathcal{O}(\frac{1}{r^2})\ .\nn
\end{align}
Here, $n'^\sigma=(0,\hat{x}')$ is normal to $r'=R$ surface. At $r'=R$ we can approximate the current by its expression for $r'>R$ given in \eqref{jj}. Let us use $\int_{r'=R} d^4x' = \int d^4x' \ \delta(r'-R)$. We will use the $\delta^4(x'-x_i(\tau))$ term in \eqref{jj} to perform the integral over $d^4x'$. 
\begin{align}
q^\sigma A^{*\text{bulk}}_\sigma(x)
&=-\frac{i}{8\pi^2 r}\  \int d\tau \sum_{j=1}^{2n}\frac{\delta(\tau-\eta_jT)}{|\p_\tau r'|\ r'(\tau)}\ e_j\ [(\vec{V}_j \tau +\vec{d}_j).\vec {V}_{j}]\ \log{(x^2-2x.V_j\tau -2x.d_j)}+\mathcal{O}(\frac{1}{r^2})\ .\nn
\end{align}
Here  $\p_\tau r'$ factor appears in the denominator as $\delta(r'-R) = \frac{\delta(\tau-\eta_jT)}{|\p_\tau r'|}$ and in our parametrisation $\tau=\eta_jT$ corresponds to $r'(\eta_jT)=R$. Thus we get 
\begin{align}
q^\sigma A^{*\text{bulk}}_\sigma(x)
&=-\frac{i}{8\pi^2 r}\   \sum_{j=1}^{2n} \eta_je_j\  \log{(x^2-2\eta_jx.V_jT -2x.d_j)}+\mathcal{O}(\frac{1}{r^2})\ .\nn
\end{align}
Expanding above expression around $\mathcal{I}^+$ 
\begin{align}
q^\sigma A^{*\text{bulk}}_\sigma(x)
&=-\frac{i}{8\pi^2 r}\ \Big[  \sum_{j=1}^{2n} \eta_je_j\  \log{(-2ur)}+\frac{1}{u}\sum_{j=1}^{2n} \eta_je_j\  (q.d_j+\eta_jq.V_j\ T)\ \Big]+\mathcal{O}(\frac{1}{u^2})\ \label{1byu bulk}
\end{align}
The first term vanishes due to conservation of charge and from the second term we get
\be 
A^{*[1,1]}_\sigma(x)|_{\text{bulk}}
=-\frac{i}{8\pi^2 }\ \sum_{j=1}^{2n} \eta_je_j\  ( \ d_{j\sigma}+\eta_jV_{j\sigma}\ T\ )\ .\nn
\ee
Adding the asymptotic contribution in \eqref{12} to above expression we can write down the full $\frac{1}{ur}$-term
\begin{align}
A^{*[1,1]}_\mu(\hat{x})
& = \frac{i}{8\pi^2}  \ \sum_{j=1}^{2n}\Big[\frac{\eta_je_jV_{j\mu}}{V_j.q}q.d_j-\eta_je_jd_{j\mu}\Big]\ .\label{11}
\end{align}
In above expression, we have the full coefficient of the $\frac{1}{ru}$ mode of the Feynman radiative solution. Interestingly this coefficient is proportional to the tree level subleading soft factor. 

Next we will argue that presence of $\frac{1}{ur}$-mode is a general feature of the quantum theory. In the previous subsection we have seen that the $\frac{\log u}{r}$ mode is related to the leading soft mode of the quantum gauge field. On the same lines, let us check if the $\frac{1}{ur}$-term is related to the (tree level) subleading soft insertion. We will extend the calculation of the previous subsection to subleading order, so we start with the second term in \eqref{d7}. We do not write the leading order term to avoid clutter.
\begin{align}
\p_u\A^{[1]+}_\mu(u,\hat{x})&= \frac{i}{8\pi^2}\frac{1}{(u-i\epsilon)}\int_0^\infty d\omega\ \p_{\omega}[\omega \tilde{A}_\mu(\omega,\hat{x})] \ e^{-i\omega (u-i\epsilon)},\nn\\
&= -\frac{1}{8\pi^2}\frac{1}{(u-i\epsilon)^2} \Big[ \p_{\omega}[\omega \tilde{A}_\mu(\omega,\hat{x})] \ e^{-i\omega (u-i\epsilon)}\Big]^\infty_0+\frac{1}{8\pi^2}\frac{1}{(u-i\epsilon)^2}\int_0^\infty d\omega\ \p^2_{\omega}[\omega \tilde{A}_\mu(\omega,\hat{x})] \ e^{-i\omega (u-i\epsilon)}.\nn\\
\end{align}
Let us consider the first term in above expression. We get 
\begin{align}
\p_u\A^{[1]+}_\mu(u,\hat{x})&=\ \frac{1}{8\pi^2} \frac{1}{(u-i\epsilon)^2} \lim_{\omega\rightarrow 0}\p_\omega[\omega \tilde{A}_\mu(\omega,\hat{x})] +...\nn  
\end{align}
The $\lim_{\omega\rightarrow 0}\p_{\omega}[\omega \tilde{A}_\mu(\omega,\hat{x})]$ operator isolates the $\omega^0$-mode in the gauge field. Integrating above expression we get
\begin{align}
\A^{[1]+}_\mu(u,\hat{x})&=-\frac{1}{8\pi^2}\ [ \frac{1}{u} +i\pi\delta(u)]\ \lim_{\omega\rightarrow 0}\p_\omega[\omega \tilde{A}_\mu(\omega,\hat{x})] +...\nn  
\end{align}
Thus we see that the tree level subleading soft mode indeed gives rise to a $\frac{1}{u}$-mode at large $u$. 
Let us evaluate its coefficient using tree level subleading soft theorem \cite{sub1,sub2}.
$$<\text{out}|\lim_{\omega\rightarrow 0^+} \p_\omega[\omega a^{(\pm)}(\omega,\hat{x})]\ S|\text{in}>\ =\sum_je_j\frac{\epsilon^{(\pm)}_\mu q_\nu}{p_j.q}\mathcal{J}_{j}^{\mu\nu} <\text{out}|S|\text{in}>\ .$$
$\mathcal{J}_{j}^{\mu\nu}$ is the angular momentum operator. Hence we have
\begin{align}
 <\text{out}|\A^{[1,1]}_\mu(\hat{x})S|\text{in}>\ & = \frac{i}{8\pi^2}  \ \big[\epsilon^+_\mu\epsilon^-_\nu+\epsilon^-_\mu\epsilon^+_\nu\big] \ \sum_{j=1}^{2n}\frac{e_j q_\lambda}{p_j.q}\ \mathcal{J}_{j}^{\nu\lambda} <\text{out}|S|\text{in}>\ ,\nn\\
&\ = \frac{i}{8\pi^2}  \ \sum_{j=1}^{2n}\frac{e_jq^\nu}{p_j.q}\ \mathcal{J}_{j\mu\nu} <\text{out}|S|\text{in}>.
\end{align}
 In the case of point particles, $\mathcal{J}_{j}^{\mu\nu}=i\eta_j(x_j^\mu p_j^\nu -p_j^\mu x_j^\nu)=i\eta_j(d_j^\mu p_j^\nu -p_j^\mu d_j^\nu)$. Substituting in above expression, we see that it matches with the expression in \eqref{11} that was derived from the Feynman solution. Here we derived the coefficient of the $\frac{1}{ru}$ mode in $A_\sigma$ using the tree level subleading soft mode. This immediately tell us that this mode will get modified when we go to higher orders in $e$. This mode is also expected to get corrected if the matter has internal spin and also in presence of non-minimal couplings.  

To summarise in this section, we discussed how the $\frac{\log u}{r}$ mode in \eqref{AFP} is controlled by the leading soft mode and the $\frac{1}{ur}$ mode in \eqref{AFP} is controlled by the (tree level) subleading soft mode. We expect this to hold for all $\frac{1}{u^n}$ modes in \eqref{AFP}; these $\frac{1}{u^n}$ modes should be related to the ${\omega}^{n-1}$ soft modes respectively. This also tells us that though we obtained \eqref{AFP} for a toy example the presence of $\log u$ and $\frac{1}{u^n}$ modes is a general feature of QED. 

\section{Effect of long range forces on asymptotic trajectories}\label{5}
Let us turn back to the problem of scattering of $n$ charged particles. At large distances, the electromagnetic force present between the scattering particles falls off as $\Oo(\frac{1}{r^2})$ and gives rise to logarithmic correction to the straight line trajectory at late times. In this section we will obtain the the explicit form of this logarithmic correction. 

We need to find the leading order term in the asymptotic electromagnetic field strength using \eqref{Afeyn}. This leading $\Oo(\frac{1}{r^2})$-mode has an imaginary piece. Thus the corrected equation of trajectory will also have an imaginary piece! This should be compared with the Faddeev-Kulish dressing of scalar fields under electromagnetic force. The logarithmic correction to the trajectory of a particle is in one to one correspondence with the logarithmic dressing of the scalar field \cite{FK} as we will discuss below.

The equation of trajectory of $j^{th}$  outgoing particle is given by :
\begin{align}
m_j\frac{\p^2x_j^\mu}{\p \tau^2}=e_j\ F^{\mu\nu}(x_j(\tau))\ V_{j\nu} .\label{eot}
\end{align}
Here, we need to find the field strength generated  at the position of the outgoing particle i.e. at $x=x_j(\tau)$ using \eqref{Afeyn}. It will get contribution from other particles that interact with the $j^{th}$ particle. Since the outgoing $j^{th}$ particle approaches $\mathcal{I}^+_+$ asymptotically, we need to evaluate \eqref{Afeyn} around $u\rightarrow\infty$. It is important to note some subtle points. As seen from \eqref{Afeyn}, the contribution from the first line to the field around $u\rightarrow\infty$ contains contribution only from outgoing particles. While the second line of \eqref{Afeyn} contains contribution from both incoming and outgoing particles. The leading order field at large $\tau$ is given by
\begin{align}
F_{\mu\nu}(x_j(\tau))|_{\mathcal{I}^+_+}
&=\frac{1}{4\pi \tau^2}\sum_{\substack{i=n+1,\\ i\neq j}} ^{2n}e_i \frac{ (V_{i\mu}V_{j\nu}-V_{j\mu}V_{i\nu}) }{[(V_i.V_j)^2-1]^{3/2}} \nn\\
+&\frac{i}{8\pi^2 \tau^2}\sum_{\substack{i=1,\\ i\neq j}} ^{2n}\eta_ie_i \frac{ (V_{i\mu}V_{j\nu}-V_{j\mu}V_{i\nu}) }{[(V_i.V_j)^2-1]^{3/2}} \Big[\log\frac{-V_i.V_j+\sqrt{(V_i.V_j)^2-1}}{-V_i.V_j-\sqrt{(V_i.V_j)^2-1}}+2V_i.V_j\ \sqrt{(V_i.V_j)^2-1}\Big]+\mathcal{O}(\frac{1}{\tau^3}) .\label{F}
\end{align}
Substituting \eqref{F} in \eqref{eot}, the leading order correction to the asymptotic trajectories of the particles are 
\be x^\mu_j= V_j^\mu \ \tau +(c^\mu_j+i\C_j^\mu) \log \tau + d_j+\mathcal{O}(\frac{1}{\tau}),\nn\ee
where we get (for outgoing particles)
\begin{align}
c_j^\mu&=\frac{1}{4\pi} \sum_{\substack{i=n+1,\\ i\neq j}} ^{2n} e_i e_j\frac{ (V_{i\mu}+V_{j\mu}V_{i}.V_j) }{[(V_i.V_j)^2-1]^{3/2}},\nn\\
\C_j^\mu&=\frac{1}{8\pi^2 }\sum_{\substack{i=1,\\ i\neq j}} ^{2n}\eta_ie_ie_j\frac{  (V_{i\mu}+V_{j\mu}V_{i}.V_j) }{[(V_i.V_j)^2-1]^{3/2}} \Big[\log\frac{-V_i.V_j+\sqrt{(V_i.V_j)^2-1}}{-V_i.V_j-\sqrt{(V_i.V_j)^2-1}}+2V_i.V_j\ \sqrt{(V_i.V_j)^2-1}\Big] .\label{ca}
\end{align}
Let us compare the logarithmic correction in the trajectory with the logarithmic dressing of the scalar field \cite{FK}. $c_j^\mu$ is related to the $\Phi$ term in the dressing of scalar field as given in eq (11) of \cite{FK} while $\C_j^\mu$ is related to the \textbf{R} term in the dressing of scalar field as given in eq (10) of \cite{FK}.
%
%
For $j^{th}$ incoming particle, the corresponding terms are given by 
\begin{align}
c_j^\mu&=-\frac{1}{4\pi} \sum_{\substack{i=1,\\ i\neq j}} ^n e_i e_j\frac{ (V_{i\mu}+V_{j\mu}V_{i}.V_j) }{[(V_i.V_j)^2-1]^{3/2}}\nn\\
\C_j^\mu&=\frac{1}{8\pi^2 }\sum_{\substack{i=1,\\ i\neq j}} ^n\sum_{\substack{i=1,\\ i\neq j}} ^{2n}\eta_ie_ie_j\frac{  (V_{i\mu}+V_{j\mu}V_{i}.V_j) }{[(V_i.V_j)^2-1]^{3/2}} \Big[\log\frac{-V_i.V_j+\sqrt{(V_i.V_j)^2-1}}{-V_i.V_j-\sqrt{(V_i.V_j)^2-1}}+2V_i.V_j\ \sqrt{(V_i.V_j)^2-1}\Big] .
\end{align}
It should be noted that in the first term, the sum over $i$ includes only the incoming particles. 

For conciseness, let us define $C^\mu_j = c^\mu_j+i\C^\mu_j$ so that 
\be x^\mu_j= V_j^\mu \ \tau +C^\mu_j \log \tau + d_j+\mathcal{O}(\frac{1}{\tau}). \label{x1}\ee
Because of this $\mathcal{O}(e^2)$ correction to the asymptotic trajectories, the current given in \eqref{jj} also gets corrected.
\be
j^{\text{asym}}_\sigma(x') &=& \int d\tau\ \Big[ \sum_{j=n+1}^{2n}e_j \big[V_{j\sigma}+\frac{C_{j\mu}}{\tau}\big]\  \delta^4(x'-x_j)\ \Theta(\tau-T)\nn\\
&&+\sum_{j=1}^{n}e_j \big[V_{j\sigma}+\frac{C_{j\mu}}{\tau}\big]\  \delta^4(x'-x_j)\ \Theta(-\tau-T)\Big].\label{j1}
\ee
Next we will find the electromagnetic field generated by above current.

\section{Radiative field at $\mathcal{O}(e^3)$ with Feynman propagator}\label{6}
In this section we will obtain the asymptotic radiative field keeping the leading order effect of long range electromagnetic force acting on the particles.  As a result of this long range force, a particle continues to accelerate at late times and this gives rise to new modes in the asymptotic field at $\mathcal{O}(e^3)$. Using \eqref{j1} we get
\begin{align}
A_\sigma(x)
&=\int_T^\infty d\tau\ \sum_{j=n+1}^{2n} G(x,x_j)\  e_j \big[V_{j\sigma}+\frac{C_{j\sigma}}{\tau}\big]\ +\int^{-T}_{-\infty} d\tau\ \sum_{j=1}^{n} G(x,x_j)\  e_j \big[V_{j\sigma}+\frac{C_{j\sigma}}{\tau}\big]\ \nn\\
&+\int_{r'<R} d^4x'\ G(x,x')\  j^{\text{bulk}}_\sigma(x')\  .\nn
\end{align}
We recall the expression of the Feynman propagator
\be
G(x;x')
&=& \frac{1}{4\pi^2 } \Big[\pi\delta_+(\ (x-x')^2)+\pi\delta_-(\ (x-x')^2)+\frac{i}{(x-x')^2}\Big]\label{fp1}.
\ee

Let us first write down the contribution from the first two terms of \eqref{fp1}. As disscussed in the beginning of section 2, these terms get contribution only from asymptotic sources and it suffices to use \eqref{j1}. We cannot solve the $\delta$-function condition exactly because of the logarithmic correction. We solve it perturbatively in Appendix \ref{pert} and quote the solution to the delta function constraint from \eqref{tau1}
\begin{align}
\tau_1^\pm&=-V_i.(x-d_i)\mp\big[\ (V_i.x-V_i.d_i)^2 +(x-d_i)^2\ -2(x-d_i).C_i\log\tau_0^\pm\ \big]^{1/2}.\label{t}
\end{align}
Here, $\tau_0^\pm$ is the zeroth order solution given in \eqref{tau00}. Hence we get
\begin{align}
A^{+}_\sigma(x)+A^{-}_\sigma(x)
&=\frac{1}{4\pi }\int d\tau\  \sum_{i=n+1}^{2n}[\delta_+(\ (x-x')^2)+\delta_+(\ (x-x')^2)\ ]\  e_i \big[V_{i\sigma}+\frac{C_{j\sigma}}{\tau}\big]\    \Theta(\tau-T)\ \ + \ \ \text{in}.\nn\\
&=\frac{1}{4\pi }\int d\tau\  \sum_{i=n+1}^{2n}\frac{\delta(\tau-\tau_1^+)+\delta(\tau-\tau_1^-)}{|2\tau+2V_i.(x-d_i)+\frac{2}{\tau}C_i.(x-d_i)|}\  e_i \big[V_{i\sigma}+\frac{C_{j\sigma}}{\tau}\big]\    \Theta(\tau-T)\ \ + \ \ \text{in}.
\nn\end{align}
%
%
We have not written the contribution if the incoming particles explicitly. Above expression is vaild only to $\mathcal{O}(e^3)$. Expanding the roots in \eqref{t} to $\mathcal{O}(e^3)$, we have :
\begin{align}
\tau^\pm_1|_{\mathcal{I}^+}&=-V_i.(x-d_i)\mp\big[\ (V_i.x-V_i.d_i)^2 +(x-d_i)^2\ -2\ \big]^{1/2}\pm(x-d_i).C_i\frac{\log\tau_0^\pm}{X}.\nn
\end{align}
Hence we get
\begin{align}
A^{+}_\sigma(x)+A^{-}_\sigma(x)
&=\frac{1}{8\pi } \sum_{i=n+1}^{2n}  \frac{\Theta(u-T)\ e_i }{X} \Big[\ V_{i\sigma} \big[1+\frac{(x-d_i).C_i}{X^2}\log\tau^+_0+\frac{(x-d_i).C_i}{X\tau^+_0}\big]+\frac{C_{i\sigma}}{\tau^+_0}\Big]\nn\\
&+\frac{1}{8\pi } \sum_{i=1}^{n}  \frac{\Theta(-u-T)\ e_i }{X} \Big[\ V_{i\sigma} \big[1+\frac{(x-d_i).C_i}{X^2}\log\tau^+_0+\frac{(x-d_i).C_i}{X\tau^+_0}\big]+\frac{C_{i\sigma}}{\tau^+_0}\Big]\ \nn\\
&+\frac{1}{8\pi } \sum_{i=n+1}^{2n}  \frac{e_i }{X} \Big[\ V_{i\sigma} \big[1+\frac{(x-d_i).C_i}{X^2}\log\tau^-_0-\frac{(x-d_i).C_i}{X\tau^-_0}\big]+\frac{C_{i\sigma}}{\tau^-_0}\Big]\ ,\label{A01}\\
&\text{where }X=[\ (V_i.x-V_i.d_i)^2 +(x-d_i)^2\ ]^{1/2}. \nn
\end{align} 
We can study the expansion of above expression around $\mathcal{I}^+$. Using \eqref{tau00} and \eqref{logt} we have
\begin{align}
&\big[A^{+}_\sigma(x)+A^{-}_\sigma(x)\big]|_{\mathcal{I}^+}\nn\\
&=\sum_{\substack{n=1,\\ m=-\infty,\\ m<n. }}^\infty  A_\mu^{[n,-m]}(\hat{x})  \ \frac{u^m}{ r^n}\   + \ \log u \sum_{\substack{n=2,m=0,\\ m<n-1}}^\infty  A_{1 \mu\ }^{[n,-m]}(\hat{x})  \ \frac{u^m}{ r^n}\   + \ \log r \sum_{\substack{n=2,m=0,\\ m<n-1}}^\infty  A_{2\mu}^{[n,-m]}(\hat{x})  \ \frac{u^m}{ r^n}+...\ .\label{+-e3}
\end{align}
Above expression should be compared with \eqref{+e} and \eqref{-e}. The logarithmic modes present in above expression appear only at $\mathcal{O}(e^3)$ and are a direct consequence of the long range electromagnetic forces present between the scattering particles. These modes are absent in \eqref{+e} and \eqref{-e}. '...' denote terms that fall off faster than any power law.
Let us turn to the contribution from the third term of \eqref{fp1}.
\begin{align}
A^*_\sigma(x)=\frac{i}{4\pi^2}\int d^4x'\  \frac{ j_\sigma(x')}{(x-x')^2}\  .\label{b1}
\end{align}
We write down the  asymptotic part using \eqref{j1}. The integral needs to be regulated with an IR cutoff  '$R$'.
\begin{align}
A^{*\text{asym}}_\sigma(x)
&=\frac{i}{4\pi^2 }\int_T^R d\tau \sum_{j=n+1}^{2n} \frac{e_j[V_{j\sigma}+\frac{C_{j\sigma}}{\tau}]}{(x-V_j\tau-d_j-C_j\log\tau)^2}+\frac{i}{4\pi^2 }\int^{-T}_{-R} d\tau \sum_{j=1}^{n}\frac{e_j[V_{j\sigma}+\frac{C_{j\sigma}}{\tau}]}{(x-V_j\tau-d_j-C_j\log\tau)^2},\nn\\
&=\frac{i}{4\pi^2 }\int_T^R d\tau \sum_{j=n+1}^{2n}\Big[ \frac{e_j[V_{j\sigma}+\frac{C_{j\sigma}}{\tau}]}{(x-V_j\tau-d_j)^2}+\frac{2(x-d_j).C_j}{(x-V_j\tau-d_j)^4}\log\tau\Big] \ \ + \ \ \text{in}\nn .
\end{align}
In above expression we have not written the contribution of the incoming particles explicitly to avoid clutter. Let us rewrite it as follows
\begin{align}
A^{*\text{asym}}_\sigma(x)
&=\frac{i}{4\pi^2 }\int_T^R d\tau \sum_{j=n+1}^{2n}\Big[ e_j\frac{[V_{j\sigma}+\frac{C_{j\sigma}}{\tau}]}{\tau_0^--\tau_0^+}\big[\frac{1}{\tau-\tau_0^+}-\frac{1}{\tau-\tau_0^-}\big]\ +\ e_jV_{j\sigma}\frac{2(x-d_j).C_j\ \log\tau}{(\tau-\tau_0^-)^2(\tau-\tau_0^+)^2}\Big] \ \ + \ \ \text{in}\label{a4} .
\end{align}
$\tau^\pm_0$ given in \eqref{tau00} are the solutions to the equation $(x-V_i\tau-d_i)^2=0$.  These integrals have been discussed in Appendix \ref{6A}. The final expression of \eqref{a4} is given in \eqref{A**1}. To this expression, we add the contribution of \eqref{A01} to get
\be
A_\sigma(x)&=&\frac{i}{4\pi^2 }\sum_{j=1}^{2n}  \frac{\eta_je_j V_{j\sigma}}{\tau_0^--\tau_0^+}\Big[\log\frac{1}{\tau_0^+-\eta_jT}-\log\frac{1}{\tau_0^--\eta_jT}\Big]\ \nn\\
&&+\frac{i}{4\pi^2 }\sum_{j=1}^{2n} \frac{\eta_je_jC_{j\sigma}}{(\tau_0^--\tau_0^+)}\ \big[\ \frac{1}{\tau_0^+}\log \frac{T}{\tau_0^+-\eta_jT}\ -\ \frac{1}{\tau_0^-}\log \frac{T}{\tau_0^--\eta_jT}\ \big]\nn\\
&&+\frac{i}{4\pi^2 }\sum_{j=1}^{2n}\dfrac{2\eta_je_jV_{j\sigma}(x-d_j).C_j}{(\tau_0^--\tau_0^+)^2}\log (T)\big[\frac{1}{T-\tau_0^+}+\frac{1}{T-\tau_0^-}\big]\nn\\
&&+\frac{i}{4\pi^2 }\sum_{j=1}^{2n}\dfrac{2\eta_je_jV_{j\sigma}(x-d_j).C_j}{(\tau_0^--\tau_0^+)^2}\big[\ \frac{1}{\tau_0^-}\log\frac{T}{\tau_0^--\eta_jT}+\frac{1}{\tau_0^+}\log\frac{T}{\tau_0^+-\eta_jT}\ \big]\nn\\
&&+\frac{i}{4\pi^2 }\sum_{j=1}^{2n}\dfrac{4\eta_je_jV_{j\sigma}(x-d_j).C_j}{(\tau_0^--\tau_0^+)^3}\big[-\ln\tau_0^+\ln (\tau_0^+-\eta_jT)+\ln\tau_0^-\ln (\tau_0^--\eta_jT)+\frac{1}{2}[\ln^2\tau_0^+-\ln^2\tau_0^-\big]\nn\\
&&-\frac{i}{4\pi^2 }\sum_{j=1}^{2n}\dfrac{4\eta_je_jV_{j\sigma}(x-d_j).C_j}{(\tau_0^--\tau_0^+)^3}\Big[\operatorname{Li}_2(1-\frac{\eta_jT}{\tau_0^-})-\operatorname{Li}_2(1-\frac{\eta_jT}{\tau_0^+
})\Big]\ \ \nn\\
&&+\frac{1}{8\pi } \sum_{i=n+1}^{2n}  \frac{\Theta(u-T)\ e_i }{X} \Big[\ V_{i\sigma} \big[1+\frac{(x-d_i).C_i}{X^2}\log\tau^+_0+\frac{(x-d_i).C_i}{X\tau^+_0}\big]+\frac{C_{i\sigma}}{\tau^+_0}\Big]\nn\\
&&+\frac{1}{8\pi } \sum_{i=1}^{n}  \frac{\Theta(-u-T)\ e_i }{X} \Big[\ V_{i\sigma} \big[1+\frac{(x-d_i).C_i}{X^2}\log\tau^+_0+\frac{(x-d_i).C_i}{X\tau^+_0}\big]+\frac{C_{i\sigma}}{\tau^+_0}\Big]\ \nn\\
&&+\frac{1}{8\pi } \sum_{i=n+1}^{2n}  \frac{e_i }{X} \Big[\ V_{i\sigma} \big[1+\frac{(x-d_i).C_i}{X^2}\log\tau^-_0-\frac{(x-d_i).C_i}{X\tau^-_0}\big]+\frac{C_{i\sigma}}{\tau^-_0}\Big]\ +A^{*\text{bulk}}_\sigma(x).\label{A*1}\ee
$X=[\ (V_i.x-V_i.d_i)^2 +(x-d_i)^2\ ]^{1/2}$. $A^{*\text{bulk}}_\sigma(x)$ denotes the contribution of the bulk sources to \eqref{b1}. As this term depends on the the detailed form of bulk trajectories, the exact form of this term cannot be obtained. This term has been estimated in \eqref{*bulk}.

We can finally write down the asymptotic expansion of the full Feynman solution at $\mathcal{O}(e^3)$. The asymptotic expansion of various terms in the solution has been discussed in \eqref{Aexp} of Appendix \ref{6A}. Using \eqref{+-e3}, \eqref{Aexp} and \eqref{*bulk}, we get
\begin{align}
A_\sigma(x)|_{\mathcal{I}^+}&= \ (\log r)^2 \sum_{\substack{n=2,\\ m=0,\\ m<n-1.}}^\infty  [A_{\ell_1\mu}^{[n,-m]}(\hat{x})]  \ \frac{u^m}{ r^n}+ \ \log r \sum_{\substack{n=1,\\ m=0,\\ m<n.}}^\infty  [A_{\ell_2\mu}^{[n,-m]}(\hat{x})]  \ \frac{u^m}{ r^n}+ \ (\log u)^2 \sum_{\substack{n=2,\\ m=0,\\ m<n-1. }}^\infty  [A_{\ell_3\mu}^{[n,-m]}(\hat{x})]  \ \frac{u^m}{ r^n}\ \nn\\
&+ \ \log u \sum_{\substack{n=1,\\ m=-\infty,\\ m<n. }}^\infty  [A_{\ell_4\mu}^{[n,-m]}(\hat{x})]  \ \frac{u^m}{ r^n}\ +\sum_{\substack{n=1,\\ m=-\infty,\\ m<n.}}^\infty  [A_\mu^{[n,-m]}(\hat{x})]  \ \frac{u^m}{ r^n}\     \ .\label{e3}
\end{align}

Above expression should be compared with \eqref{AFP}. It is important to study the $\Oo(e^3)$ corrections to $\frac{1}{r}$ term of $A_\sigma$ :
\begin{align}
A_\sigma(x)|_{\mathcal{I}^+}\sim\ \frac{1}{ r} \ [\ \log u + u^0 + \sum_{m=1}^\infty\frac{\log u}{u^m} + \sum_{n=1}^\infty\frac{1}{u^n} + ...\ ]\   .\label{A111}
\end{align}
This expression should be compared with its analogue at $\Oo(e)$ given in \eqref{AFP1}. Using \eqref{A*1}, it can be shown that the $\log u$ and the $u^0$ modes are not modified at $\Oo(e^3)$ . The $\frac{\log u}{u^m }$-modes are absent at $\Oo(e)$ and arise as a result of long range electromagnetic interactions between the scattering particles. It should be noted that the coefficients of the $\frac{1}{u^n}$-modes are modified at $\Oo(e^3)$. This is consistent with the fact that the $\omega^{n-1}$-soft modes for $n>0$ also gets corrected at this order.

\subsection{The $\frac{\log u}{ur}$ mode in $A_\sigma$}
In this subsection we will discuss if the $\frac{\log u}{u^m }$-modes can be related to the soft modes. Let us study the coefficient of the $\frac{\log u}{ur}$ term in Feynman radiative field $A_\sigma$. The second and fourth line of \eqref{A*1} contribute to this mode. We get 
\be
A^{[1,\frac{\log u}{u}]}_\sigma(\hat{x})
&=&-\frac{i}{8\pi^2 }\sum_{j=1}^{2n}   [\ \eta_je_jC_{j\sigma} - \eta_je_jV_{j\sigma}\dfrac{ q.C_j}{q.V_i}\ ] \label{logu by u}.\ee
Using the expression of $C_j^\mu$ from \eqref{x1} we see that the coefficient of the $\frac{\log u}{ur}$-mode is proportional to the loop level subleading soft theorem\cite{1808.03288}. This hints that this mode is related to the loop level subleading soft mode. 

Let us check if the $\frac{\log u}{u}$-term is related to the loop level subleading soft mode of the quantum gauge field. We need to start with the second term in \eqref{d7}. We do not write the leading order term in \eqref{d7} to avoid clutter.
\begin{align}
\p_u\A^{[1]+}_\mu(u,\hat{x})
&=\frac{i}{8\pi^2}\frac{1}{u}\int_0^\infty d\omega\ \p_{\omega}[\omega \tilde{A}_\mu(\omega,\hat{x})] \ e^{-i\omega (u-i\epsilon). },\nn\\
&=-\frac{1}{8\pi^2}\frac{1}{(u-i\epsilon)^2}\int_0^\infty d\omega\ \p_{\omega}[\omega \tilde{A}_\mu(\omega,\hat{x})] \ \p_\omega e^{-i\omega (u-i\epsilon)}\nn .
\end{align}
The subleading term in the soft expansion of gauge field at 1-loop is logarithmic in energy\cite{1808.03288}. We will manipulate above expression in a way that isolates this mode.
\begin{align}
\p_u[u^2\p_u\A^{[1]+}_\mu(u,\hat{x})]
&=\frac{i}{8\pi^2}\int_0^\infty d\omega\ \p_{\omega}[\omega \tilde{A}_\mu(\omega,\hat{x})] \ \p_\omega [\omega e^{-i\omega (u-i\epsilon)}],\nn\\
&=\frac{i}{8\pi^2}\Big[\omega \p_{\omega}[\omega \tilde{A}_\mu(\omega,\hat{x})]  e^{-i\omega (u-i\epsilon)}\Big]_0^\infty-\frac{i}{8\pi^2}\int_0^\infty d\omega\ \omega\p^2_{\omega}[\omega \tilde{A}_\mu(\omega,\hat{x})] \  e^{-i\omega (u-i\epsilon)}\nn\\
\end{align}
The first term goes to 0 at 1-loop level. From the second term we get 
\begin{align}
\p_u[u^2\p_u\A^{[1]+}_\mu(u,\hat{x})]
&=\frac{1}{8\pi^2}\frac{1}{(u-i\epsilon)}\int_0^\infty d\omega\ \omega\p^2_{\omega}[\omega \tilde{A}_\mu(\omega,\hat{x})] \  \p_\omega e^{-i\omega (u-i\epsilon)}\nn\\
&=\frac{1}{8\pi^2}\frac{1}{(u-i\epsilon)}\Big[ \omega\p^2_{\omega}[\omega \tilde{A}_\mu(\omega,\hat{x})e^{-i\omega (u-i\epsilon)}\Big]_0^\infty+...\nn
\end{align}
Hence we get 
\begin{align}
\p_u\A^{[1]+}_\mu(u,\hat{x})&=-\frac{1}{8\pi^2}\frac{1}{u^2} \ [\log u +i\pi\Theta(u)]\ \lim_{\omega\rightarrow 0}[\omega\p^2_{\omega}[\omega \tilde{A}_\mu(\omega,\hat{x})] +...\nn  
\end{align}
\begin{align}
\A^{[1]+}_\mu(u,\hat{x})&=\frac{1}{8\pi^2} \ [\frac{\log u}{u} +i\pi\frac{\Theta(u)}{u}]\ \lim_{\omega\rightarrow 0}[\omega\p^2_{\omega}[\omega \tilde{A}_\mu(\omega,\hat{x})] +...\nn  
\end{align}
The $\lim_{\omega\rightarrow 0}[\omega\p^2_{\omega}[\omega \tilde{A}_\mu(\omega,\hat{x})]$ operator isolates the soft log $\omega$ mode in the gauge field. Thus we see that the soft log $\omega$ mode gives rise to $\f{\log u}{u}$ and $\frac{1}{u}$ modes at large $u$.
The $\frac{\Theta(u)}{u}$ mode is responsible for the tail memory effect discussed in \cite{log mem,log mem em}.

Here we are interested in the $\frac{\log u}{u}$ mode. The coefficient of this mode is given by
\begin{align}
\A^{[1,\frac{\log u}{u}]}_\mu(u,\hat{x})& = \frac{1}{8\pi^2}  \lim_{\omega\rightarrow 0^+} \omega\p^2_\omega[\omega \tilde{A}_\mu(\omega,\hat{x})]\ ,\nn\\
& =\frac{1}{8\pi^2}\lim_{\omega\rightarrow 0^+} \  \omega\p^2_\omega[\omega a_\mu(\omega,\hat{x})] \ .\nn
\end{align}
Let us evaluate the coefficient using loop level subleading soft theorem which is given by \cite{1808.03288}
$$<\text{out}|\lim_{\omega\rightarrow 0^+} \omega\p^2_\omega[\omega a^{(\pm)}(\omega,\hat{x})]\ S|\text{in}>\ =\sum_{j=1}^{2n}\Big[\frac{\eta_je_jp_j^\nu}{p_j.q}q.C_j-\eta_je_jC_j^\nu\Big] <\text{out}|S|\text{in}>\ .$$
$C_j$'s take value as given in \eqref{x1}. Hence we get
\begin{align}
 <\text{out}|\A^{[1,\frac{\log u}{u}]}_\mu(\hat{x})S|\text{in}>\ 
&\ = \frac{i}{8\pi^2}  \ \big[\epsilon^+_\mu\epsilon^-_\nu+\epsilon^-_\mu\epsilon^+_\nu\big] \ \sum_{j=1}^{2n}\Big[\frac{\eta_je_jp_j^\nu}{p_j.q}q.C_j-\eta_je_jC_j^\nu\Big]\  <\text{out}|\ S|\text{in}>\ ,\nn\\
&\ = \frac{i}{8\pi^2}  \ \sum_{j=1}^{2n}\Big[\frac{\eta_je_jp_{j\mu}}{p_j.q}q.C_j-\eta_je_jC_{j\mu}\Big]\  <\text{out}|S|\text{in}>.
\end{align}
Thus the S matrix element matches with \eqref{logu by u} that we obtained using the Feynman solution. 

We conclude that the $\frac{\log u}{ur}$-term is related to the loop level subleading soft insertion. It is likely that the $\frac{\log u}{u^m r }$-modes for $m>1$ are related to the $\omega^m\log\omega$ soft modes that also appear at 1-loop order. But these soft modes are not expected to be universal. This hints that the quantum gauge field will in general contain $\frac{\log u}{u^m r }$-modes such that the $m=1$ mode is universal whereas the $m>1$ modes are not.

\subsection{Expectation at higher orders in $e$}
Let us first discuss what kind of new modes will appear in $A_\mu$ at $\Oo(e^5)$. We have seen that at $\mathcal{O} (e^3)$,  new modes arise in the radiative field due to acceleration of the charged particles under the long range electromagnetic force. This radiation backreacts on the particles. When we go to higher orders in $e$ we need to include the effect of this backreaction.
It can shown using \eqref{A*1} that the asymptotic forces $(F)$ acting on a scattered particle take following form at $\mathcal{O} (e^3)$, 
$$ F(\tau) =e\ \sum_{m=2}^\infty\frac{c_m}{\tau^m}+e^3\ \sum_{n=3}^\infty d_n\frac{\log\tau}{\tau^n}.$$
The leading order $\frac{1}{\tau^2}$-term is the $\mathcal{O}(e)$ term that we had studied earlier. It gives rise to logarithmic correction to the trajectory we had discussed in \eqref{x1}. This term does not get corrected at $\mathcal{O} (e^3)$. Similar to the analysis of \cite{2007.03627} we see that the next order correction to the asymptotic trajectories of the particles is of following form 
\be x^\mu_i=V^\mu_i \tau +C^\mu_i \log \tau + d_i+\tilde{C}_{i\sigma}\frac{\log\tau}{\tau}. \nn \ee
Next we will repeat the steps we followed earlier. First we need to find the correction to the current given in \eqref{j1} and  then find the resultant field using the Feynman propagator. We expect that the $\frac{1}{r}$ term of $A_\sigma$ takes following form 
\begin{align}
A_\sigma(x)\ \sim\  \frac{e}{r}\ [\ \log u +u^0+\frac{1}{u}+\frac{1}{u^2}+...\ ] + \frac{e^3}{r}\frac{\log u}{u}\ [\ 1+\frac{1}{u}+...\ ]+ \frac{e^5}{r}\frac{(\log u)^2}{u^2}\ [\ 1+\frac{1}{u}+...\ ] .\nn
\end{align}
$\frac{(\log u)^2}{u^2r}$ is absent in the retarded solution\cite{2007.03627}. We also think that this mode should controlled by the $\omega(\log\omega)^2$ soft mode that appears at 2-loop order\cite{Bis}. Since this soft mode is universal, the $\frac{(\log u)^2}{u^2r}$-mode should also be universal.

In general we anticipate that the $\frac{1}{r}$ term of the radiative field takes following form 
\begin{align}
A_\sigma(x)\ \sim\ \frac{1}{r}\ [\  \log u +  \ u^0+ \sum_{\substack{m=0,\\ n=1,\\ n\geq m.}}^\infty \frac{( \log u)^m}{u^n}\ ]\ .\
\end{align}
The first two terms are universal and exact at $\Oo(e)$. Here the terms in the summation are such that the $m^{th}$ term appears at $\mathcal{O}(e^{2m+1})$. We also surmise that the $ \frac{( \log u)^m}{u^m}$-modes are controlled by the $\omega^{m-1}(\log\omega)^m$ universal soft modes that are expected to appear at $m^{th}$-loop order. All the modes except the $\log u$ mode obey Ashtekar-Struebel conditions\cite{AS}.

\section{The new asymptotic conservation law} \label{7}
We discussed that new modes arise in the asymptotic field at $\mathcal{O}(e^3)$ as a result of the long range interactions between scattering particles. In this section we will obtain an asymptotic conservation equation obeyed by such modes.

Let us first discuss the $Q_1$ conservation law derived in \cite{2007.03627} for the retarded solution. In \cite{2007.03627} it was shown that $F_{rA}$ (which denotes the '$rA$'-component of the field strength) has following expansion near $u \rightarrow -\infty$.
\be
F^{[\text{ret}]}_{rA}|_{\mathcal{I}^+_-} = \frac{1}{r^2}\ [\ u\ F^{(u)}_{rA}(\hat{x})\ +\ \log u\ F^{(\log u)}_{rA}(\hat{x})\ + ... ] +\mathcal{O}( \frac{1}{r^3})\ . \ee
Similarly around the past null infinity near $v \rightarrow \infty$ 
\be
F^{[\text{ret}]}_{rA}|_{\mathcal{I}^-_+} =  \frac{\log r}{r^2}\ [v^0 \ F^{(\log r)}_{rA}(\hat{x})\ +...]+\mathcal{O}( \frac{1}{r^2})\ .\ee
The $Q_1$ conservation law derived in \cite{2007.03627} is given as follows
\begin{align}
F^{(\log u)}_{rA}(\hat{x})\ |_{\mathcal{I}^+_-}&=F^{(\log r)}_{rA}(-\hat{x})\ |_{\mathcal{I}^-_+}.\label{conQ1}
\end{align}
The future charge is defined as $Q^+_1=\int d^2z \ Y^A(\hat{x})\ F^{[\log u]}_{rA}(\hat{x})|_{\mathcal{I}^+_-}$ and the past charge is $Q^-_1=\int d^2z \ Y^A(-\hat{x})\ F^{[\log r]}_{rA}(-\hat{x})|_{\mathcal{I}^-_+}$. These charges are $\mathcal{O}(e^{3})$. In \cite{1903.09133,1912.10229}, the authors started with these 'classical' charges and showed that upon quantisation these charges reproduce the full $\log\omega$-soft theorem\cite{1808.03288} including the purely quantum modes.

We aim to find the analogue of \eqref{conQ1} that is obeyed by the Feynman solution. 
$F_{rA}$ calculated using \eqref{A*1} takes following form 
\begin{align}
F_{rA}|_{\mathcal{I}^+_-} = \frac{\log r}{r^2}\ F_{rA}^{[\log r]}(\hat{x}) +\frac{1}{r^2}\ [\ u\log u \ F^{[u\log u]}_{rA}(\hat{x})+ (\log u)^2\ F^{[(\log u)^2]}_{rA}(\hat{x})+ u\ F^{[u]}_{rA}(\hat{x})\ +\ \log u\ F^{[\log u]}_{rA}(\hat{x}) + ... ]\ . \end{align}
Similarly around the past null infinity we have : 
\begin{align}
F_{rA}|_{\mathcal{I}^-_+} =  \frac{\log r}{r^2}\ F^{[\log r]}_{rA}(\hat{x})+\frac{1}{r^2}\ [\ v\log v \ F^{[v\log v]}_{rA}(\hat{x})+ (\log v)^2\ F^{[(\log v)^2]}_{rA}(\hat{x})+ v\ F^{[v]}_{rA}(\hat{x})\ +\ \log v\ F^{[\log v]}_{rA}(\hat{x}) + ... ] \ .\end{align}
In this section we will derive the conservation equation obeyed by these modes.
\begin{align}
[F^{[\log u]}_{rA}(\hat{x})-F^{[\log r]}_{rA}(\hat{x})]\ |_{\mathcal{I}^+_-}&=[-F^{[\log v]}_{rA}(-\hat{x})+F^{[\log r]}_{rA}(-\hat{x})]\ |_{\mathcal{I}^-_+}.
\end{align}
So that the future charge can be defined as $\tilde{Q}^+_1=\int d^2z \ Y^A(\hat{x})\ [F^{[\log u]}_{rA}(\hat{x})-F^{[\log r]}_{rA}(\hat{x})]|_{\mathcal{I}^+_-}$ and the past charge by $\tilde{Q}^-_1=\int d^2z \ Y^A(-\hat{x})\ [-F^{[\log v]}_{rA}(-\hat{x})+F^{[\log r]}_{rA}(-\hat{x})]|_{\mathcal{I}^-_+}$. 

\subsection{Modes at Future null infinity}
Let us find the full $\frac{\log u}{r^2}$-mode of $A_\sigma$ at the future null infinity.  We need to expand all the terms in the Feynman solution given in \eqref{A*1} around $\mathcal{I}^+$. First we turn to the seventh and eighth lines of \eqref{A*1}. Using \eqref{t1f}, we have $\log\tau_0|_{\mathcal{I}^+}\sim \log u+\mathcal{O}(u^0)$. 
Using \eqref{q}, we get $X=-rq.V_i+\mathcal{O}(r^0)$.  Substituting the limiting value of $X$, we can read off the coefficient of the $\mathcal{O}(\frac{\log u}{r^2})$ term in the seventh and eighth lines of \eqref{A*1} :
\begin{align}
&-\frac{1}{8\pi } \sum_{i=n+1}^{2n} \Theta(u-T)\  e_i \ V_{i\sigma} \ \frac{q.C_i}{(q.V_i)^3}-\frac{1}{8\pi } \sum_{i=1}^{n} \Theta(-u-T)  e_i \ V_{i\sigma} \ \frac{q.C_i}{(q.V_i)^3}
\end{align}
It can be shown that above term contains the full soft factor correspoding to the $\log \omega$ term\cite{1808.03288}.

There are many more terms that contribute to the $\frac{\log u}{r^2}$-mode. These remaining terms are actually not related for the $\log\omega$ soft theorem\cite{1808.03288}. We will list them. From the first line of \eqref{A*1}, using \eqref{d3} we get 
\begin{align}
\frac{i}{8\pi^2 }\ \sum_{j=1}^{2n} \eta_j\frac{e_j V_{j\sigma}}{(V_j.q)^2}\  [\ \frac{q.d_j}{(V_j.q)}+V_j.d_j\ ].
\end{align}
From the second line of \eqref{A*1}, we get using \eqref{d1} and \eqref{d3}
\begin{align}
-\frac{i}{16\pi^2 }\sum_{j=1}^{2n}\eta_j e_jC_{j\sigma}\  [\ \frac{1}{(V_j.q)^2}-1\ ] .
\end{align}
The third line of \eqref{A*1} does not have a log $u$ term. From the fourth line of \eqref{A*1}, we get using \eqref{d1} and \eqref{d3}
\begin{align}
-\frac{i}{8\pi^2 }\sum_{j=1}^{2n}\eta_j \frac{e_jV_{j\sigma}}{V_j.q}\ \Big[\ q.C_j  [- \frac{3}{2(V_j.q)^2}-\frac{V_j^0}{(V_j.q)}+\frac{1}{2}\ ] +C_j^0\ \Big].
\end{align}
Using \eqref{d3}, the fifth line of \eqref{A*1} gives 
\be
-\frac{i}{8\pi^2 }\sum_{j=1}^{2n}\eta_je_jV_{j\sigma}\dfrac{q.C_j}{(q.V_j)^3}\ \ln |q.V_j|.
\ee
As shown in Appendix \ref{6A}, the sixth line of \eqref{A*1} and $A^{*\text{bulk}}_\sigma$ do not have any logarithmic modes. We have the full coefficient of the $\frac{\log u}{r^2}$ term.
\begin{align}
A^{[\log u/r^2]}_\sigma|_{\mathcal{I}^+_-}&=-\frac{1}{8\pi } \sum_{i=1}^{n}  e_i \ V_{i\sigma} \ \frac{q.C_i}{(q.V_i)^3}+\frac{i}{8\pi^2 }\ \sum_{j=1}^{2n} \eta_j\frac{e_j V_{j\sigma}}{(V_j.q)^2}\  [\ \frac{q.d_j}{(V_j.q)}+V_j.d_j\ ]\nn\\
&-\frac{i}{16\pi^2 }\sum_{j=1}^{2n}\eta_je_j C_{j\sigma}\  [\ \frac{1}{(V_j.q)^2}-1\ ]-\frac{i}{8\pi^2 }\sum_{j=1}^{2n}\eta_je_jV_{j\sigma}\dfrac{q.C_j}{(q.V_j)^3}\ \ln |q.V_j|\nn\\
&-\frac{i}{8\pi^2 }\sum_{j=1}^{2n}\eta_j \frac{e_jV_{j\sigma}}{V_j.q}\ \Big[\ q.C_j  [-\frac{3}{2(V_j.q)^2}-\frac{V_j^0}{(V_j.q)}+\frac{1}{2}\ ] +C_j^0\ \Big].\label{loguf}
\end{align}

Next we need to write down the coefficient of the $\frac{\log r}{r^2}$-term in $A_\sigma$. 
From the first line of \eqref{A*1}, using \eqref{d3} we get
\begin{align}
-\frac{i}{8\pi^2 }\ \sum_{j=1}^{2n} \eta_j\frac{e_j V_{j\sigma}}{(V_j.q)^2}\  [\ \frac{q.d_j}{(V_j.q)}+V_j.d_j\ ].
\end{align}
From the second line of \eqref{A*1}, using \eqref{d2} and \eqref{d3} we get
\begin{align}
\frac{i}{16\pi^2 }\sum_{j=1}^{2n}\eta_j \frac{e_jC_{j\sigma}}{(q.V_j)^2} .
\end{align}
The third line of \eqref{A*1} does not have a log $r$ term. From the fourth line of \eqref{A*1}, using \eqref{d2} and \eqref{d3} we get
\begin{align}
\frac{i}{16\pi^2 }\sum_{j=1}^{2n}\eta_j \frac{e_jV_{j\sigma}}{(V_j.q)^3}\ q.C_j   .
\end{align}
The fifth line of \eqref{A*1} contributes as follows
\be
-\frac{i}{8\pi^2 }\sum_{j=1}^{2n}\eta_je_jV_{j\sigma}\dfrac{q.C_j}{(q.V_j)^3}\ \ln( 2|q.V_j|).
\ee
Substituting $X=-rq.V_i+\mathcal{O}(r^0)$ in the eighth line of \eqref{A*1}, we get
\begin{align}
&-\frac{1}{8\pi } \sum_{i=n+1}^{2n}\  e_i \ V_{i\sigma} \ \frac{q.C_i}{(q.V_i)^3}.
\end{align}
We have the full coefficient of the $\frac{\log r}{r^2}$ term.
\begin{align}
A^{[\log r/r^2]}_\sigma(x)|_{\mathcal{I}^+}
&=-\frac{1}{8\pi } \sum_{i=n+1}^{2n}\  e_i \ V_{i\sigma} \ \frac{q.C_i}{(q.V_i)^3}-\frac{i}{8\pi^2 }\ \sum_{j=1}^{2n} \eta_j\frac{e_j V_{j\sigma}}{(V_j.q)^2}\  [\ \frac{q.d_i}{(V_i.q)}+V_j.d_j\ ]\nn\\
&+\frac{i}{16\pi^2 }\sum_{j=1}^{2n}\eta_j \frac{e_jC_{j\sigma}}{(q.V_j)^2} +\frac{i}{16\pi^2 }\sum_{j=1}^{2n}\eta_j \frac{e_jV_{j\sigma}}{(V_j.q)^3}\ q.C_j-\frac{i}{8\pi^2 }\sum_{j=1}^{2n}\eta_je_jV_{j\sigma}\dfrac{q.C_j}{(q.V_j)^3}\ \ln(2 |q.V_j|).\label{logrf}
\end{align}

\subsection{Modes at Past null infinity}
Next we need to derive the field configuration at past null infinity and then compare the two expressions. Analogous to \eqref{A*1}, around $\mathcal{I}^-$ we have
\be
A_\sigma(x)&=&\frac{i}{4\pi^2 }\sum_{j=1}^{2n}  \frac{\eta_je_j V_{j\sigma}}{\tau_0^--\tau_0^+}\Big[\log\frac{1}{\tau_0^+-T}-\log\frac{1}{\tau_0^--T}\Big]\ \nn\\
&&+\frac{i}{4\pi^2 }\sum_{j=1}^{2n} \frac{\eta_je_jC_{j\sigma}}{(\tau_0^--\tau_0^+)}\ \big[\ \frac{1}{\tau_0^+}\log \frac{T}{\tau_0^+-T}\ -\ \frac{1}{\tau_0^-}\log \frac{T}{\tau_0^--T}\ \big]\nn\\
&&+\frac{i}{4\pi^2 }\sum_{j=1}^{2n}\dfrac{2\eta_je_jV_{j\sigma}(x-d_j).C_j}{(\tau_0^--\tau_0^+)^2}\log T\big[\frac{1}{T-\tau_0^+}+\frac{1}{T-\tau_0^-}\big]\nn\\
&&+\frac{i}{4\pi^2 }\sum_{j=1}^{2n}\dfrac{2\eta_je_jV_{j\sigma}(x-d_j).C_j}{(\tau_0^--\tau_0^+)^2}\big[\ \frac{1}{\tau_0^-}\log\frac{T}{\tau_0^--T}+\frac{1}{\tau_0^+}\log\frac{T}{\tau_0^+-T}\ \big]\nn\\
&&+\frac{i}{4\pi^2 }\sum_{j=1}^{2n}\dfrac{4\eta_je_jV_{j\sigma}(x-d_j).C_j}{(\tau_0^--\tau_0^+)^3}\big[-\ln\tau_0^+\ln (\tau_0^+-T)+\ln\tau_0^-\ln (\tau_0^--T)+\frac{1}{2}[\ln^2\tau_0^+-\ln^2\tau_0^-\big]\nn\\
&&-\frac{i}{4\pi^2 }\sum_{j=1}^{2n}\dfrac{4\eta_je_jV_{j\sigma}(x-d_j).C_j}{(\tau_0^--\tau_0^+)^3}\Big[\operatorname{Li}_2(-\frac{T-\tau_0^-}{\tau_0^-})-\operatorname{Li}_2(-\frac{T-\tau_0^+}{\tau_0^+
})\Big]\ \ \nn\\
&&+\frac{1}{8\pi } \sum_{i=n+1}^{2n}  \frac{\Theta(v-T)\ e_i }{X} \Big[\ V_{i\sigma} \big[1+\frac{(x-d_i).C_i}{X^2}\log\tau^-_0-\frac{(x-d_i).C_i}{X\tau^-_0}\big]+\frac{C_{i\sigma}}{\tau^-_0}\Big]\nn\\
&&+\frac{1}{8\pi } \sum_{i=1}^{n}  \frac{\Theta(-v-T)\ e_i }{X} \Big[\ V_{i\sigma} \big[1+\frac{(x-d_i).C_i}{X^2}\log\tau^-_0-\frac{(x-d_i).C_i}{X\tau^-_0}\big]+\frac{C_{i\sigma}}{\tau^-_0}\Big]\ \nn\\
&&+\frac{1}{8\pi } \sum_{i=1}^{n}  \frac{e_i }{X} \Big[\ V_{i\sigma} \big[1+\frac{(x-d_i).C_i}{X^2}\log\tau^+_0+\frac{(x-d_i).C_i}{X\tau^+_0}\big]+\frac{C_{i\sigma}}{\tau^+_0}\Big]\ .\label{A*2}
\ee
Here, $X=[\ (V_i.x-V_i.d_i)^2 +(x-d_i)^2\ ]^{1/2}. $

Let us write down the full coefficient of the $\frac{\log v}{r^2}$-mode. We need to take the limit $r\rightarrow\infty$ with $v=t+r$ finite. In this co-ordinate system, 4 dimensional spacetime point can be parametrised as :
\be x^\mu = r\bar{q}^\mu + v t^\mu,\ \ \  \bar{q}^\mu=(-1,\hat{x}), \ \ \  t^\mu=(1,\vec{0}).\label{qbar}\ee
$\bar{q}^\mu$ is a null vector. 
From the first line of \eqref{A*2}, we get using \eqref{d6}
\begin{align}
\frac{i}{8\pi^2 }\ \sum_{j=1}^{2n} \eta_j\frac{e_j V_{j\sigma}}{(V_j.\q)^2}\  [\ \frac{\q.d_j}{(V_j.\q)}+V_j.d_j\ ].
\end{align}
From the second line of \eqref{A*2}, using \eqref{d5} and \eqref{d6} we get
\begin{align}
-\frac{i}{16\pi^2 }\sum_{j=1}^{2n}\eta_j e_jC_{j\sigma}\   [\ \frac{1}{(V_j.\q)^2}-1\ ] .
\end{align}
The third line of \eqref{A*2} does not have a log $v$ term. From the fourth line of \eqref{A*2}, using \eqref{d5} and \eqref{d6} we get
\begin{align}
-\frac{i}{8\pi^2 }\sum_{j=1}^{2n}\eta_j \frac{e_jV_{j\sigma}}{V_j.\q}\ \Big[\ \q.C_j  [\ \frac{V_j^0}{(V_j.\q)}-\frac{3}{2(\q.V_i)^2}+\frac{1}{2}\ ] -C_j^0\ \Big].
\end{align}
The fifth line of \eqref{A*2} gives 
\be
&-&\frac{i}{8\pi^2 }\sum_{j=1}^{2n}\eta_je_jV_{j\sigma}\dfrac{\q.C_j}{(\q.V_j)^3}\ \ln |\q.V_j|\ .
\ee
From seventh line of \eqref{A*2} we get 
\be  \frac{1}{8\pi } \sum_{i=n+1}^{2n} \Theta(v-T)\  e_i \ V_{i\sigma} \ \frac{\bar{q}.C_i}{(\bar{q}.V_i)^3}+\frac{1}{8\pi } \sum_{i=1}^{n} \Theta(-v-T)  e_i \ V_{i\sigma} \ \frac{\bar{q}.C_i}{(\bar{q}.V_i)^3}.\ee
Hence we have 
\begin{align}
A^{[\log v/r^2]}_\sigma(x)
&=+\frac{1}{8\pi } \sum_{i=n+1}^{2n} e_i \ V_{i\sigma} \ \frac{\bar{q}.C_i}{(\bar{q}.V_i)^3}+\frac{i}{8\pi^2 }\ \sum_{j=1}^{2n} \eta_j\frac{e_j V_{j\sigma}}{(V_j.\q)^2}\  [\ \frac{\q.d_j}{(V_j.\q)}+V_j.d_j\ ]\nn\\
&-\frac{i}{16\pi^2 }\sum_{j=1}^{2n}\eta_j e_jC_{j\sigma}\  [\ \frac{1}{(V_j.\q)^2}-1\ ]-\frac{i}{8\pi^2 }\sum_{j=1}^{2n}\eta_je_jV_{j\sigma}\dfrac{\q.C_j}{(\q.V_j)^3}\ \ln |\q.V_j|\nn\\
&-\frac{i}{8\pi^2 }\sum_{j=1}^{2n}\eta_j \frac{e_jV_{j\sigma}}{V_j.\q}\ \Big[\ \q.C_j  [-\frac{3}{2(V_j.\q)^2}+\frac{V_j^0}{(V_j.\q)}+\frac{1}{2}\ ] -C_j^0\ \Big].\label{logvp}
\end{align}
Next we turn to the $\frac{\log r}{r^2}$-mode.
From the first line of \eqref{A*2}, we get
\begin{align}
-\frac{i}{8\pi^2 }\ \sum_{j=1}^{2n} \eta_j\frac{e_j V_{j\sigma}}{(V_j.\q)^2}\  [\ \frac{\q.d_j}{(V_j.\q)}+V_j.d_j\ ].
\end{align}
From the second line of \eqref{A*2}, we get
\begin{align}
\frac{i}{16\pi^2 }\sum_{j=1}^{2n}\eta_j e_jC_{j\sigma}\  \frac{1}{(V_j.\q)^2}.
\end{align}
The third line of \eqref{A*2} does not have a log $r$ term. From the fourth line of \eqref{A*2}, we get
\begin{align}
\frac{i}{16\pi^2 }\sum_{j=1}^{2n}\eta_j \frac{e_jV_{j\sigma}}{(V_j.\q)^3}\  \q.C_j .
\end{align}
The fifth line of \eqref{A*2} gives 
\be
-\frac{i}{8\pi^2 }\sum_{j=1}^{2n}\eta_je_jV_{j\sigma}\dfrac{\q.C_j}{(\q.V_j)^3}\ \ln |2\q.V_j|\ .
\ee
We get following contribution from the last line of \eqref{A*2}
\be \frac{1}{8\pi } \sum_{i=1}^{n}  e_i \ V_{i\sigma} \ \frac{\bar{q}.C_i}{(\bar{q}.V_i)^3}.\ee
The total coefficient is
\begin{align}
A^{[\log r/r^2]}_\sigma(x)|_{\mathcal{I}^-}
&=\frac{1}{8\pi } \sum_{i=1}^{n} e_i \ V_{i\sigma} \ \frac{\bar{q}.C_i}{(\bar{q}.V_i)^3}-\frac{i}{8\pi^2 }\ \sum_{j=1}^{2n} \eta_j\frac{e_j V_{j\sigma}}{(V_j.\q)^2}\  [\ \frac{\q.d_j}{(V_j.\q)}+V_j.d_j\ ]\nn\\
&+\frac{i}{16\pi^2 }\sum_{j=1}^{2n}\eta_j e_jC_{j\sigma}\  \frac{1}{(V_j.\q)^2}+\frac{i}{16\pi^2 }\sum_{j=1}^{2n}\eta_j \frac{e_jV_{j\sigma}}{(V_j.\q)^3}\  \q.C_j -\frac{i}{8\pi^2 }\sum_{j=1}^{2n}\eta_je_jV_{j\sigma}\dfrac{\q.C_j}{(\q.V_j)^3}\ \ln |2\q.V_j|\ .\label{logrp}
\end{align}

Thus, from \eqref{loguf}, \eqref{logrf}, \eqref{logvp} and \eqref{logrp} we can indeed check that following modes are equal under antipodal idenfication.
\begin{align}
[A^{[\log u/r^2]}_\sigma(\hat{x})-A^{[\log r/r^2]}_\sigma(\hat{x})]\ |_{\mathcal{I}^+_-}&=[A^{[\log v/r^2]}_\sigma(-\hat{x})-A^{[\log r/r^2]}_\sigma(-\hat{x})]\ |_{\mathcal{I}^-_+}.\label{1lup}
\end{align}
Using co-ordinate transformation, it can be shown that the Feynman solution obeys following conservation equation :
\begin{align}
[F^{[\log u/r^2]}_{rA}(\hat{x})-F^{[\log r/r^2]}_{rA}(\hat{x})]\ |_{\mathcal{I}^+_-}&=[-F^{[\log v/r^2]}_{rA}(-\hat{x})+F^{[\log r/r^2]}_{rA}(-\hat{x})]\ |_{\mathcal{I}^-_+}.\label{1loop}
\end{align}
Compared to \eqref{1lup}, the RHS of above expression has extra minus sign as it has an extra factor of $\p_A\q^\mu$ due to co-ordinate transformation. Finally we have derived the $\tilde{Q}_1$-conservation equation such that the future charge is defined by $\tilde{Q}^+_1=\int d^2z \ Y^A(\hat{x})\ [F^{[\log u/r^2]}_{rA}(\hat{x})-F^{[\log r/r^2]}_{rA}(\hat{x})]|_{\mathcal{I}^+_-}$ and the past charge by $\tilde{Q}^-_1=\int d^2z \ Y^A(-\hat{x})\ [-F^{[\log v/r^2]}_{rA}(-\hat{x})+F^{[\log r/r^2]}_{rA}(-\hat{x})]|_{\mathcal{I}^-_+}$. 

Let us state some important observations. $\tilde{Q}^+_1$ gets contribution from \eqref{loguf} and \eqref{logrf} and it is seen that it contains terms that are not related to the $\log\omega$ mode. 
We suspect that such irrelevant terms would cancel from the Ward identity and the $\tilde{Q}_1$-charge will reproduce the full $\log\omega$ soft theorem\cite{1808.03288}. Nonetheless using $\tilde{Q}^+_1$-conservation law given in \eqref{1loop} is not a satisfactory way of identifying the asymptotic charge. \cite{1903.09133,1912.10229} have a different prescription to define the asymptotic charges. The classical law given in \eqref{conQ1} is used to define asymptotic charges, these charges are then quantised and the corresponding Ward identity is shown to be equivalent to the full $\log{\omega}$ soft photon theorem. But it should be noted that \eqref{conQ1} itself is violated in presence of Feynamn boundary condition. Hence, using either of the conservation equations given in \eqref{conQ1} and \eqref{1loop} to define asymptotic charges is not completely satisfactory.  It would be useful to have a first principles-based construction of these asymptotic charges via asymptotic phase space techniques.

\section{Summary}
In this paper we have obtained the radiative field produced by scattering of $n$ charged point particles using Feynman propagator. This problem is unphysical but the Feynman radiative solution so derived is useful to illustrate interesting apects of the quantum gauge field.

We showed in \eqref{AFP1} that the $\frac{1}{r}$-term in the Feynman solution at $\mathcal{O}(e)$ has following behaviour 
\begin{align}
A_\mu(x)|_{\mathcal{I}^+}=\frac{1}{ r} \ [\ \log u + u^0 + \sum_{n=1}^\infty\frac{1}{u^n} + ...\ ] \ .\nn
\end{align}
Here, '...' denote terms that fall off faster than any power law in $u$. 
The log $u$ and the $\frac{1}{u^n}$-modes are purely quantum modes; they are absent in the retarded solution in \eqref{AA1}. It should be noted that the $\log u$ mode violates the Ashtekar-Struebel fall offs for the radiative field\cite{AS} that ensure the existence of a well defined symplectic form. We leave the investigation of this issue to the future. 

The $\log u$ mode is controlled by the leading soft mode\cite{1903.09133,1912.10229}. Extending this idea, we  showed that the $\frac{1}{u}$-mode is related to the tree level subleading soft mode in the quantum gauge field.  This is an interesting result and it hints that the $\frac{1}{u^n}$ modes for $m>1$ should similarly be related to the ${\omega}^{n-1}$ soft modes  respectively. Hence we expect that the presence of $\log u$ and $\frac{1}{u^n}$ modes is a general feature of QED. 

New modes are expected to appear in the radiative field as we go to higher orders in the coupling. Including the effect of long range electromagnetic force on the scattering particles, we obtained the Feynman solution upto $\Oo(e^3)$  in \eqref{A*1}. The $\frac{1}{r}$-term takes following form
 \begin{align}
A_\sigma(x)|_{\mathcal{I}^+}\ \sim\ \frac{1}{ r} \ [\ \log u + u^0 + \sum_{m=1}^\infty\frac{\log u}{u^m} + \sum_{n=1}^\infty\frac{1}{u^n} + ...\ ]\ .\nn
\end{align}
The $\frac{\log u}{u^n}$-modes in the Feynman solution appear at $\mathcal{O}(e^3)$. These modes go to 0 as $u\rightarrow\pm\infty$ and do not violate the Ashtekar-Struebel fall offs \cite{AS}. It should also be noted that the $\frac{\log u}{u}$-mode is absent in the retarded solution\cite{2007.03627}. We studied the coefficient of the $\frac{\log u}{u}$ quantum mode and showed that this mode is related to the loop level soft $\log{\omega}$-mode. As the $\log{\omega}$-mode derived in \cite{1808.03288} is universal, the $\frac{\log u}{u}$-mode should also be universally present in the quantum gauge field.

We expect that new $\frac{(\log u)^m}{u^m}$-modes would appear in the Feynman solution at $\mathcal{O}(e^{2m+1})$ respectively such that they are related to the $\omega^{m-1}(\log\omega)^m$ universal soft modes
The $\frac{1}{r}$-term of the radiative field speculatively takes following form 
\begin{align}
A_\sigma(x)|_{\mathcal{I}^+}\ \sim\  \frac{1}{r}\ [\ e \log u + e \ u^0+ \sum_{\substack{m=0,\\ n=1,\\ n\geq m.}}^\infty \frac{( \log u)^m}{u^n}\ ]\  ,\nn
\end{align}
such that the $m^{th}$ term in the summation appears at $\mathcal{O}(e^{2m+1})$.  Hence it is expected that all the modes in the Feynman solution except the $\log u$ mode should preserve the Ashtekar-Struebel fall offs.

In section \ref{7}, we turned to the asymptotic conservation equation obeyed by certain $\Oo(e^3)$ logarithmic modes in the Feynman solution. This equation has been derived in \eqref{1loop} and relates the difference in the coefficients of the $\frac{\log u}{r^2}$ and $\frac{\log r}{r^2}$ modes in $F_{rA}$ at $\mathcal{I}^+_-$ to the difference in the coefficients of the $\frac{\log r}{r^2}$ and $\frac{\log v}{r^2}$ in $F_{rA}$ at $\mathcal{I}^-_+$. It should be possible to prove \eqref{1loop} in general by following analysis of \cite{soft inf} albeit with Feynman boundary condition. It is expected that the corresponding charges $\tilde{Q}_1$ should reproduce the $\log{\omega}$ soft photon theorem derived in \cite{1808.03288}. These questions need to be pursued in the future.


\section{Acknowledgements}
I am extremely thankful to Nabamita Banerjee and Alok Laddha for numerous discussions. I am thankful to Prof Ashoke Sen for his comments. I thank the participants of the program 'Recent Developments in S-matrix
theory' for discussion and I also thank International Centre for
Theoretical Sciences (ICTS) for hosting this online program (code:
ICTS/rdst2020/07). I am deeply grateful to my family for their constant support. Finally I thank the people of India for their enduring help to basic sciences.

\appendix
\section{Feynman propagator}\label{FP}
We are mostly familiar with the momentum space form of the Feynman propagator. With normalisation such that $\Box G=-\delta(x-x')$, it takes following form 
\be
G(x,x')
&=&\  \int \frac{d^4p}{(2\pi)^{4}}\   \frac{e^{ip.(x-x')}}{p^2-i\epsilon}.\nn\\
\ee
We can perform the $p^0$ intgral according to the given presciption and get
\be
G(x,x')
&=&\  i\int \frac{d^3p}{(2\pi)^{3}}\  \frac{1}{2\omega}\ [ e^{ip.(x-x')} \Theta(t-t') +e^{-ip.(x-x')} \Theta(t'-t) ]\nn\\
\ee
Here $\omega =|\vec{p}|$. Let us work in a frame s.t. $\hat{x}-\hat{x}' \ || \ p_z$-axis. We have $d^3p=\omega^2d\omega \ d\phi\ d\cos\theta$ and the integral over $d\phi$ gives $2\pi$.
\begin{align}
&G(x,x')
=&\ i\int_{-1}^1 d\cos\theta \frac{\omega d\omega}{2(2\pi)^{2}}\   [ e^{-i\omega(t-t'-|x-x'|\cos\theta)} \Theta(t-t') +e^{i\omega(t-t'-|x-x'|\cos\theta)} \Theta(t'-t) ].
\end{align}
Performing the $\cos\theta$ integral 
\begin{align}
&G(x,x')= \frac{1}{|x-x'|}\int_0^\infty \frac{ d\omega}{2(2\pi)^{2}}\   \Big[ e^{-i\omega(t-t')}[e^{i\omega |x-x'|}-e^{-i\omega |x-x'|}] \Theta(t-t') -e^{i\omega(t-t')}[e^{-i\omega |x-x'|}-e^{i\omega |x-x'|}] \Theta(t'-t) \Big].
\end{align}
In the last line, we recall that $\omega>0$, the integral is the standard Fourier transform integral 
\be \int d\omega\ e^{-i\omega u}\ \Theta(\omega) = -\frac{i}{u}+\pi\delta(u). \label{FT} \ee
Hence we get  
\be
G(x;x')
&=& \frac{1}{8\pi^2 } \frac{ \Theta(t-t')}{|x-x'|} \Big[-\frac{i}{t-t'-|x-x'|}+\pi\delta(t-t'-|x-x'|)+\frac{i}{t-t'+|x-x'|}\Big]\nn\\
&&+ \frac{1}{8\pi^2 } \frac{\Theta(t'-t)}{|x-x'|} \Big[-\frac{i}{t-t'-|x-x'|}+\pi\delta(t-t'+|x-x'|)+\frac{i}{t-t'+|x-x'|}\Big].\nn
\ee
We can rewrite above expression as 
\be
G(x;x')&=& \frac{1}{4\pi^2 } \Big[\frac{i}{(x-x')^2}+\pi\delta_+(\ (x-x')^2)+\pi\delta_-(\ (x-x')^2)\Big]\label{fpp}
\ee
Here, the subscript '+' denotes the retarded root $t-t'-|x-x'|=0$ while the subscript '-' denotes the advanced root $t-t'+|x-x'|=0$.

\section{Perturbative solution}\label{pert}

The Green function for d'Alembertian operator is $\delta([x-x']^2)$. We will find the solution of this delta function perturbatively in coupling $e$. Here, $x'^\mu(\tau)$ is the equation of trajectory that gets corrected as we go to higher orders in $e$. We will write down the perturbative solution for $\tau$.

At zeroth order, we have free particles :
$$ x'^\mu_i= V_i^\mu \tau + d_i.$$
Hence, the root of delta function $\delta([x-x']^2)$ is given by :
\begin{align}
\tau_0^\pm&=-V_i.(x-d_i)\mp\big[\ (V_i.x-V_i.d_i)^2 +(x-d_i)^2\ \big]^{1/2}.\label{tau00}
\end{align}
$\tau_0^+$ satisifies retarded boundary condition while $\tau_0^-$ satisifies advanced boundary condition.
Let us study above expression in the limit $r\rightarrow \infty$ with $u$ finite. Thus, around $\mathcal{I}^+$, using \eqref{q} we get  :
\be \tau^+_0|_{\mathcal{I}^+}=\frac{u+q.d_i}{|q.V_i|}+\mathcal{O}(\frac{1}{r}),\ \ \tau_0^-|_{\mathcal{I}^+}= 2r{|q.V_i|}+\mathcal{O}(r^0).\label{tau0f}\ee
Now we take $r\rightarrow \infty$ limit of \eqref{tau00} keeping $v$ finite, using \eqref{qbar}, we get :
\begin{align}
\tau^+_0|_{\mathcal{I}^-}&=-2r\ V_i.\bar{q}   +\mathcal{O}(r^0),\ \ \tau_0^-|_{\mathcal{I}^-}= \frac{v-\bar{q}.d_i}{\bar{q}.V_i}+\mathcal{O}(\frac{1}{r}).\nn
\end{align}

Next we include the leading order effect of long range electromagnetic force. We know that the first order correction to the trajectory is given by \eqref{x1} :
$$ x'^\mu_i=V^\mu_i \ \tau +C^\mu_i \log \tau + d_i.$$
Using the corrected trajectory, the solution of delta function $\delta(|x-x'|^2)$ is given by :
\be\tau^2+2\tau V_i.(x-d_i) -(x-d_i)^2=-2(x-d_i).C_i\log\tau+C_i^2(\log \tau)^2. \nn\ee
Here we have used the fact that $V_i.C_i=0$. Noting that $C_i^\mu$ is $\mathcal{O}(e^2)$, the RHS of above equation can be treated as a perturbation. Hence we substitute the zeroth order solution \eqref{tau00} in RHS of above equation; it leads to following equation for $\tau$ :
\be\tau^2+2\tau V_i.(x-d_i) -(x-d_i)^2=-2(x-d_i).C_i\log\tau_0^\pm. \nn\ee
We ignored the $C_i^2$ term as it is $\mathcal{O}(e^4)$. Now, above equation is just a quadratic equation in $\tau$ and the solution is given by :
\begin{align}
\tau_1^\pm&=-V_i.(x-d_i)\mp\big[\ (V_i.x-V_i.d_i)^2 +(x-d_i)^2\ -2(x-d_i).C_i\log\tau_0^\pm\ \big]^{1/2}.\label{tau1}
\end{align}
We have used a subscript 1 to denote that it includes the first order perturbative effects. We can expand the squareroot to $\mathcal{O}(e^2)$ : 
\begin{align}
\tau_1^\pm&=-V_i.(x-d_i)\mp\big[\ (V_i.x-V_i.d_i)^2 +(x-d_i)^2\big]^{1/2}\ \pm\frac{(x-d_i).C_i}{X}\log\tau_0^\pm\ .\label{tau11}
\end{align}
Here, we have defined $X=[\ (V_i.x-V_i.d_i)^2 +(x-d_i)^2\ ]^{1/2}$ and $\tau_0^\pm$ are given in \eqref{tau00}. 
 Expanding around $\mathcal{I}^+$, we get :
\begin{align}
\tau_1^+|_{\mathcal{I}^+}&=-\frac{u+q.d_i}{q.V_i}- \ \frac{q.C_i}{q.V_i} \ \log \frac{(-u)}{q.V_i}+\mathcal{O}(\frac{1}{u}),\nn\\
\tau_1^-|_{\mathcal{I}^+}&=-2r{q.V_i}+ \ \frac{q.C_i}{q.V_i} \ \log r+\mathcal{O}(r^0).\label{t1f}
\end{align}
Expanding \eqref{tau11} around $\mathcal{I}^-$, we get :
\begin{align}
\tau^+_1|_{\mathcal{I}^-}&=-2r\ V_i.\bar{q}-  \frac{\bar{q}.C_i}{V_i.\bar{q}} \ \log r+\mathcal{O}(r^0),\nn\\
\tau_1^-|_{\mathcal{I}^-}&=\frac{v-\bar{q}.d_i}{\bar{q}.V_i}+ \ \frac{\bar{q}.C_i}{\bar{q}.V_i} \ \log \frac{v}{q.V_i}+\mathcal{O}(\frac{1}{v}). \label{t1p}
\end{align}

\section{Integral in section \ref{6}}\label{6A}

Let us first write down the indefinite integral given in \eqref{a4}.
\be &&\int \frac{dx\ \log x}{(x-\tau_0^-)^2(x-\tau_0^+)^2}\nn\\
&=&\dfrac{2}{(\tau_0^--\tau_0^+)^3}\big[\ln\tau_0^+\ln (x-\tau_0^+)-\ln\tau_0^-\ln (x-\tau_0^-)\big]\nn\\
&&+\dfrac{2}{(\tau_0^--\tau_0^+)^3}\Big[\operatorname{Li}_2\left(-\frac{x-\tau_0^-}{\tau_0^-}\right)-\operatorname{Li}_2\left(-\frac{x-\tau_0^+}{\tau_0^+}\right)\Big]-\frac{\ln x }{(\tau_0^+-\tau_0^-)^2}\big[\frac{1}{\left(x-\tau_0^+\right)}+\frac{1}{\left(x-\tau_0^-\right)}\big]\nn\\
&&-\frac{1}{(\tau_0^+-\tau_0^-)^2}\big[\ \frac{1}{\tau_0^-}\log\frac{x}{(x-\tau_0^-)}+\frac{1}{\tau_0^+}\log\frac{x}{(x-\tau_0^+)}\ \big]
\ee
Above integral is to be integrated from $T$ to $R$ for outgoing particles. Let us consider the upper limit and show that the divergent terms (in the $R\rightarrow\infty$ limit) indeed cancel and also find if there is any finite contribution.
\be
&&\dfrac{2}{(\tau_0^--\tau_0^+)^3}\Big[\ln\tau_0^+\ln R-\ln\tau_0^-\ln R+\operatorname{Li}_2\left(-\frac{R-\tau_0^-}{\tau_0^-}\right)-\operatorname{Li}_2\left(-\frac{R-\tau_0^+}{\tau_0^+}\right)\Big]+\mathcal{O}(\frac{\ln R }{R}).\label{a5}
\ee
Let us use following property of the dilogarithm function \cite{dilog}.
$$\operatorname{Li}_2(x) = -\frac{\pi^2}{6}-\frac{1}{2}\log(1-x)\ [2\log(-x)-\log(1-x)]\ + \ \operatorname{Li}_2(\frac{1}{1-x}).$$
Thus we have
\be
\operatorname{Li}_2\left(-\frac{R-\tau_0^+}{\tau_0^+}\right)&=& -\frac{\pi^2}{6}-\frac{1}{2}\log(\frac{R}{\tau_0^+})\ [2\log(\frac{R}{\tau_0^+}-1)-\log(\frac{R}{\tau_0^+})]\ + \ \operatorname{Li}_2(\frac{\tau_0^+}{R})\nn\\
&&=-\frac{\pi^2}{6}-\frac{1}{2}\log^2(\frac{R}{\tau_0^+})\ + \ \mathcal{O}(\frac{1}{R}).\nn
\ee
Hence \eqref{a5} is equal to 
\be
&&\dfrac{1}{(\tau_0^--\tau_0^+)^3}[\ln^2\tau_0^+-\ln^2\tau_0^-]+\mathcal{O}(\frac{\ln R }{R}).\nn
\ee
Now we can write down the result of the definite integral.
\be &&\int_T^R  \frac{dx\ \log x}{(x-\tau_0^-)^2(x-\tau_0^+)^2}\nn\\
&=&\dfrac{2}{(\tau_0^--\tau_0^+)^3}\big[\ln\tau_0^+\ln\frac{1}{\tau_0^+-T}-\ln\tau_0^-\ln \frac{1}{\tau_0^--T}+\frac{1}{2}[\ln^2\tau_0^+-\ln^2\tau_0^-]\ \big]\nn\\
&&-\dfrac{2}{(\tau_0^--\tau_0^+)^3}\Big[\operatorname{Li}_2\left(-\frac{T-\tau_0^-}{\tau_0^-}\right)-\operatorname{Li}_2\left(-\frac{T-\tau_0^+}{\tau_0^+}\right)\Big]+\frac{\ln T }{(\tau_0^+-\tau_0^-)^2}\big[\frac{1}{\left(T-\tau_0^+\right)}+\frac{1}{\left(T-\tau_0^-\right)}\big]\nn\\
&&+\frac{1}{(\tau_0^+-\tau_0^-)^2}\big[\ \frac{1}{\tau_0^-}\log\frac{T}{\tau_0^--T}+\frac{1}{\tau_0^+}\log\frac{T}{\tau_0^+-T}\ \big]\label{a6}
\ee
Hence we can write down the result of the both integraks in \eqref{a4}.
\be
&&A^{*\text{asym}}_\sigma(x)\nn\\
&=&\frac{i}{4\pi^2 }\sum_{j=n+1}^{2n}  \frac{e_j V_{j\sigma}}{\tau_0^--\tau_0^+}\Big[\log\frac{1}{\tau_0^+-T}-\log\frac{1}{\tau_0^--T}\Big]\ \nn\\
&&+\frac{i}{4\pi^2 }\sum_{j=n+1}^{2n} \frac{e_jC_{j\sigma}}{(\tau_0^--\tau_0^+)}\ \big[\ \frac{1}{\tau_0^+}\log \frac{T}{\tau_0^+-T}\ -\ \frac{1}{\tau_0^-}\log \frac{T}{\tau_0^--T}\ \big]\nn\\
&&+\frac{i}{4\pi^2 }\sum_{j=n+1}^{2n}\dfrac{2e_jV_{j\sigma}(x-d_j).C_j}{(\tau_0^--\tau_0^+)^2}\log T\big[\frac{1}{T-\tau_0^+}+\frac{1}{T-\tau_0^-}\big]\nn\\
&&+\frac{i}{4\pi^2 }\sum_{j=n+1}^{2n}\dfrac{2e_jV_{j\sigma}(x-d_j).C_j}{(\tau_0^--\tau_0^+)^2}\big[\ \frac{1}{\tau_0^-}\log\frac{T}{\tau_0^--T}+\frac{1}{\tau_0^+}\log\frac{T}{\tau_0^+-T}\ \big]\nn\\
&&+\frac{i}{4\pi^2 }\sum_{j=n+1}^{2n}\dfrac{4e_jV_{j\sigma}(x-d_j).C_j}{(\tau_0^--\tau_0^+)^3}\big[-\ln\tau_0^+\ln (\tau_0^+-T)+\ln\tau_0^-\ln (\tau_0^--T)+\frac{1}{2}[\ln^2\tau_0^+-\ln^2\tau_0^-]\ \big]\nn\\
&&-\frac{i}{4\pi^2 }\sum_{j=n+1}^{2n}\dfrac{4e_jV_{j\sigma}(x-d_j).C_j}{(\tau_0^--\tau_0^+)^3}\Big[\operatorname{Li}_2(-\frac{T-\tau_0^-}{\tau_0^-})-\operatorname{Li}_2(-\frac{T-\tau_0^+}{\tau_0^+
})\Big]\ \ + \ \ \text{in}.\label{A**1}\ee

Let us study the expansion of various terms in above expression. Using \eqref{tau00}, it is seen that 
\be \tau_0^+|_{\mathcal{I}^+}& \sim& u \ [\ 1 + \frac{1}{u}+\sum_{\substack{0\leq m\leq n,\\ n=1}}^\infty\frac{u^m}{r^n}\ ]. \nn\\
\tau_0^-|_{\mathcal{I}^+}& \sim& r + u+r^0u^0+\sum_{\substack{0\leq m\leq {n+1},\\ n=1}}^\infty\frac{u^m}{r^n}\ .\nn
\ee
 
\be\log \tau_0^-|_{\mathcal{I}^+}&\sim&\ \log r\ + \sum_{\substack{m,n=0,\\ m\leq n.}}^\infty \frac{u^m}{r^n}.\nn\\
\log \tau_0^+|_{\mathcal{I}^+}&\sim\ & \log u\ + \sum_ {\substack{n=0,\\ m=-\infty,\\m\leq n.}}^\infty \frac{u^m}{r^n}.\label{logt}\ee

\be
 \frac{1}{(\tau_0^--\tau_0^+)}\ \frac{1}{\tau_0^+}\log \frac{T}{(T-\tau_0^+)}\ &\sim&\frac{1}{r}\ [\mathcal{O}(1)+\log u]\ [\frac{1}{u}+\frac{1}{u^2}+...\ + \sum_{\substack{m=-\infty,\\ m\leq n,\\ n=1}}^\infty\frac{u^m}{r^n}].\nn\\
\frac{1}{(\tau_0^--\tau_0^+)} \frac{1}{\tau_0^-}\log \frac{T}{(T-\tau_0^-)}\ \big]&\sim&\frac{1}{r^2}\ [\mathcal{O}(1)+\log r]\ [1\ + \sum_{\substack{ 0<m\leq n,\\ n=1}}^\infty\frac{u^m}{r^n}].\nn\\
\frac{(x-d_j).C_j}{(\tau_0^+-\tau_0^-)^2}\big[\frac{1}{\left(T-\tau_0^+\right)}+\frac{1}{\left(T-\tau_0^-\right)}\big]&\sim&\frac{1}{r}\  [\frac{1}{u}+\frac{1}{u^2}+...\ + \sum_{\substack{ m=-\infty,\\ m\leq n,\\ n=1}}^\infty\frac{u^m}{r^n}].\nn\\
\frac{(x-d_j).C_j}{(\tau_0^+-\tau_0^-)^2}\frac{1}{\tau_0^+}\log\frac{T}{(T-\tau_0^+)}&\sim&\frac{1}{r}\ [\mathcal{O}(1)+\log u]\  [\frac{1}{u}+\frac{1}{u^2}+...\ + \sum_{\substack{ m=-\infty,\\ m\leq n,\\ n=1}}^\infty\frac{u^m}{r^n}].\nn\\
\frac{(x-d_j).C_j}{(\tau_0^+-\tau_0^-)^2}\frac{1}{\tau_0^-}\log\frac{T}{(T-\tau_0^-)}\ &\sim&\frac{1}{r^2}\ [\mathcal{O}(1)+\log r]\ [1\ + \sum_{\substack{ 0<m\leq n,\\ n=1}}^\infty\frac{u^m}{r^n}].\nn\\
\frac{(x-d_j).C_j}{(\tau_0^+-\tau_0^-)^3}\ln\tau_0^+\ln (T-\tau_0^+)&\sim&\frac{1}{r^2}\ \Big[ (\log u)^2\ [1 + \sum_{\substack{ 0\leq m\leq n,\\ n=1}}^\infty\frac{u^m}{r^n}]+[\mathcal{O}(1)+\log u]\ \big[1 + \sum_{\substack{m=-\infty,\\ m\leq n,\\ n=0}}^\infty\frac{u^m}{r^n}\big]\ \Big].\nn\\
\frac{(x-d_j).C_j}{(\tau_0^+-\tau_0^-)^3}\ln\tau_0^-\ln (T-\tau_0^-)&\sim&\frac{1}{r^2}\ [\mathcal{O}(1)+(\log r)^2+\log r]\ [1 + \sum_{\substack{ 0\leq m\leq n,\\ n=1}}^\infty\frac{u^m}{r^n}].\nn\\
\frac{(x-d_j).C_j}{(\tau_0^+-\tau_0^-)^3}\ \operatorname{Li}_2(-\frac{T-\tau_0^+}{\tau_0^+})\ \ &\sim& \frac{1}{r^2} \ \sum_{\substack {n=0,\\ m=-\infty,\\ m\leq n.}}\frac{u^m}{r^n}\ .\nn\\
\frac{(x-d_j).C_j}{(\tau_0^+-\tau_0^-)^3}\ \operatorname{Li}_2(-\frac{T-\tau_0^-}{\tau_0^-})\ \ &\sim& \frac{1}{r^2} \ \sum_{\substack {m,n=0,\\ m\leq n.}}\frac{u^m}{r^n}\ . \label{Aexp} \ee\\\\

\section{Appendix for section \ref{7}}\label{7A}
To find the coefficients of $\frac{\log u}{r^2}$ and $\frac{\log r}{r^2}$ modes in $A_\sigma$, we need to calculate some lower order terms in the asymptotic expansion of \eqref{A*1} explicitly. Here we list the asymptotic expansions of various quantities that appear in \eqref{A*1}.\\\\
\textbf{Around $\mathcal{I}^+$}\\\\
Let us start with the retarded root 
\begin{align}
\tau_0^+&=-V_i.(x-d_i)-\big[\ (V_i.x-V_i.d_i)^2 +(x-d_i)^2\ \big]^{1/2}.\nn
\end{align}
Around future null infinity, we get using \eqref{q}
\be
\tau_0^+|_{\mathcal{I}^+}&=&-V_i.x+V_i.d_i+rV_i.q\ \big[1-2\frac{uV_i^0+V_i.d_i}{rV_i.q}+\frac{(uV_i^0+V_i.d_i)^2}{r^2(V_i.q)^2}+\frac{(x-d_i)^2}{r^2(V_i.q)^2}\big]^{\frac{1}{2}}\nn\\
&=&-\frac{u+q.d_i}{(V_i.q)}-\frac{u^2}{2r(V_i.q)}-\frac{u^2V_i^0}{r(V_i.q)^2}-\frac{u^2}{2r(V_i.q)^3}+\mathcal{O}(\frac{u}{r}).
\ee
Hence we have 
\begin{align}
\frac{1}{\tau_0^+}&=-\frac{(V_i.q)}{u}\Big[1-\frac{q.d_i}{u}+\mathcal{O}(\frac{1}{u^2})-\frac{u}{r}[\frac{1}{2}+\frac{V_i^0}{(V_i.q)}+\frac{1}{2(V_i.q)^2}]+\mathcal{O}(\frac{u^0}{r})\Big]\ .\label{d1}
\end{align}

Next we turn to the advanced root.
\begin{align}
\tau_0^-&=-V_i.(x-d_i)+\big[\ (V_i.x-V_i.d_i)^2 +(x-d_i)^2\ \big]^{1/2}.\nn
\end{align}
Around future null infinity, we get using \eqref{q}
\be
\tau_0^-&=&-V_i.x+V_i.d_i-rV_i.q\ \big[1-2\frac{uV_i^0+V_i.d_i}{rV_i.q}+\frac{(uV_i^0+V_i.d_i)^2}{r^2(V_i.q)^2}+\frac{(x-d_i)^2}{r^2(V_i.q)^2}\big]^{\frac{1}{2}}\nn\\
&=&-2rV_i.q+2uV_i^0+2V_i.d_i+\frac{u+q.d_i}{(V_i.q)}+\mathcal{O}(\frac{1}{r}).\nn\\
\frac{1}{\tau_0^-}&=&-\frac{1}{2(V_i.q)r }\Big[1+\frac{uV_i^0+V_i.d_i}{r(V_i.q)}+\frac{u+q.d_i}{2r(V_i.q)^2}+\mathcal{O}(\frac{1}{r^2})\Big].\label{d2}
\ee
Also we can write down the asymptotic expansion of following term.
\be
\frac{2 }{\tau_0^--\tau_0^+ }&=&\frac{1}{r|V_i.q|}\ \big[1-2\frac{uV_i^0+V_i.d_i}{rV_i.q}+\frac{(uV_i^0+V_i.d_i)^2}{r^2(V_i.q)^2}+\frac{(x-d_i)^2}{r^2(V_i.q)^2}\big]^{-\frac{1}{2}},\nn\\
&=&-\frac{1}{r}\frac{1}{V_i.q}\ \big[1+\frac{1}{r}\frac{uV_i^0}{V_i.q}+\frac{1}{r}\frac{V_i.d_i}{V_i.q}+\frac{1}{r}\frac{u+d_i.q}{(V_i.q)^2}+\mathcal{O}(\frac{1}{r^2})\ \big].\label{d3}
\ee\\ . \\
\textbf{Around $\mathcal{I}^-$}\\\\
We start with the advanced root .
\begin{align}
\tau_0^-&=-V_i.(x-d_i)+\big[\ (V_i.x-V_i.d_i)^2 +(x-d_i)^2\ \big]^{1/2}.\nn
\end{align}
Around past null infinity, we get using \eqref{qbar}
\be
\tau_0^-&=&-V_i.x+V_i.d_i+rV_i.\bar{q}\ \big[1-2\frac{vV_i^0+V_i.d_i}{rV_i.\q}+\frac{(vV_i^0+V_i.d_i)^2}{r^2(V_i.\q)^2}+\frac{(x-d_i)^2}{r^2(V_i.\q)^2}\big]^{\frac{1}{2}}\nn\\
&=&\frac{v-\q.d_i}{(V_i.\q)}-\frac{v^2}{2r(V_i.\q)}+\frac{v^2V_i^0}{r(V_i.\q)^2}-\frac{v^2}{2r(V_i.\q)^3}+\mathcal{O}(\frac{v}{r}).
\ee
Hence we have
\be
\frac{1}{\tau_0^-}&=&\frac{(V_i.\q)}{v}\Big[1+\frac{\q.d_i}{v}+\mathcal{O}(\frac{1}{v^2})+\frac{v}{r}[\ \frac{1}{2}-\frac{V_i^0}{(V_i.\q)}+\frac{1}{2(V_i.\q)^2}\ ]+\mathcal{O}(\frac{v^0}{r})\Big].\label{d4}
\ee
Similarly for the retarded root we get
\be
\tau_0^+&=&-V_i.x+V_i.d_i-rV_i.\q\ \big[1-2\frac{vV_i^0+V_i.d_i}{rV_i.\q}+\frac{(vV_i^0+V_i.d_i)^2}{r^2(V_i.\q)^2}+\frac{(x-d_i)^2}{r^2(V_i.\q)^2}\big]^{\frac{1}{2}}\nn\\
&=&-2rV_i.\q+2vV_i^0+2V_i.d_i-\frac{v-\q.d_i}{(V_i.\q)}+\mathcal{O}(\frac{1}{r})\nn\\
\frac{1}{\tau_0^+}&=&-\frac{1}{2(V_i.\q)r }\Big[1+\frac{vV_i^0+V_i.d_i}{r(V_i.\q)}-\frac{v-\q.d_i}{2r(V_i.\q)^2}+\mathcal{O}(\frac{1}{r^2})\Big]\label{d5}
\ee
and
\be
\frac{2 }{\tau_0^--\tau_0^+ }&=&\frac{1}{rV_i.\q}\  \big[1-2\frac{vV_i^0+V_i.d_i}{rV_i.\q}+\frac{(vV_i^0+V_i.d_i)^2}{r^2(V_i.\q)^2}+\frac{(x-d_i)^2}{r^2(V_i.\q)^2}\big]^{-\frac{1}{2}},\nn\\
&=&\frac{1}{r}\frac{1}{V_i.\q}\ \big[1+\frac{1}{r}\frac{vV_i^0}{V_i.\q}+\frac{1}{r}\frac{V_i.d_i}{V_i.\q}-\frac{1}{r}\frac{v-d_i.\q}{(V_i.\q)^2}+\mathcal{O}(\frac{1}{r^2})\ \big].\label{d6}
\ee

\end{document}